\documentclass[prx,aps,showpacs,twocolumn,
amsmath,amssymb,superscriptaddress,nofootinbib,longbibliography,
nofootinbibjustification=justified,singlelinecheck=false]{revtex4-2}

\usepackage{physics}
\usepackage{braket}
\usepackage{bm}
\usepackage{graphicx}
\usepackage{booktabs}
\usepackage{pifont}
\usepackage{xcolor}
\usepackage{colortbl}
\usepackage{makecell}
\usepackage{array}
\usepackage{multirow}

\usepackage{float}

\newcolumntype{Y}{>{\centering\arraybackslash}X}
\newcommand{\cmark}{\textcolor{blue!70!black}{\ding{51}}}
\newcommand{\xmark}{\textcolor{red!70!black}{\ding{55}}}

\usepackage{lipsum}
\usepackage{hyperref} 
\usepackage{cleveref}
\usepackage{listings}
\usepackage{color}
\usepackage[x11names,svgnames]{xcolor}

\hypersetup{
    hidelinks,
    colorlinks=true,
    breaklinks=true,
    citecolor=SteelBlue,
    filecolor=LimeGreen,
    linkcolor=MediumBlue,
    urlcolor=MediumPurple
}

\usepackage[frozencache]{minted}

\usepackage[most]{tcolorbox}
\usepackage{inconsolata} 

\definecolor{codebg}{HTML}{F7F7F2}
\definecolor{codeframe}{HTML}{D8D8D0}

\definecolor{triagered}{HTML}{FCE8E6}

\newtcolorbox{movemain}{
  enhanced,
  breakable,
  colback=triagered,
  frame hidden,
  boxrule=0pt,
  arc=0pt,
  boxsep=0pt,
  left=0pt,
  right=0pt,
  top=2pt,
  bottom=2pt,
  before skip=4pt,
  after skip=4pt
}

\tcbuselibrary{minted}

\NewTCBListing{pythoncode}{ !O{} }{ 
  listing engine=minted,
  minted language=python,
  minted style=friendly,
  minted options={
    fontsize=\footnotesize,
    breaklines=true,
    autogobble,
    tabsize=4,
    xleftmargin=0pt
  },
  before upper={\let\centering\raggedright\raggedright},
  colback=codebg,
  colframe=codeframe,
  boxrule=0.5pt,
  arc=2pt,
  left=2mm,
  right=2mm,
  top=1.5mm,
  bottom=1.5mm,
  width=\linewidth,
  before skip=6pt,
  after skip=6pt,
  listing only,
  #1
}

\newcommand{\affTuebingen}{
Department of Computer Science, Faculty of Science,
University of T\"ubingen, Germany.
}

\newcommand{\affFeyer}{
Feyer, T\"ubingen, Germany.
}

\newcommand{\affOttawa}{
Nexus for Quantum Technologies, University of Ottawa, Canada.
}

\newcommand{\affJena}{
Institut für Festkörpertheorie und Optik,
Friedrich-Schiller-Universität Jena, Germany.
}
\begin{document}

\title{
{\normalfont\LARGE\textsc{esQueranto}}:\\[0.3em]
Differentiable Structured Quantum Light for Automated Scientific Discovery
}

\author{Marcello Armezzani}
\email{marcello.armezzani@uni-tuebingen.de}
\affiliation{\affTuebingen}
\affiliation{\affFeyer}

\author{Tareq Jaouni}
\affiliation{\affTuebingen}
\affiliation{\affOttawa}

\author{Pontus Lindgren}
\affiliation{\affTuebingen}

\author{S\"oren Arlt}
\affiliation{\affTuebingen}
\affiliation{\affFeyer}

\author{Mario Krenn}
\email{mario.krenn@uni-tuebingen.de}
\affiliation{\affTuebingen}

\author{Xuemei Gu}
\email{xuemei.gu@outlook.com}
\affiliation{\affJena}

\begin{abstract}
Uncovering new phenomena in nature requires experiments. For centuries, their design has been exclusively a human endeavor. Today, a new paradigm is emerging in which artificial intelligence and computational methods can design experiments themselves. Realizing this form of automated scientific discovery requires simulators that are sufficiently expressive to represent the diverse physical processes and couplings from which new experiments can be constructed. In this direction, we introduce \textsc{esQueranto},  a differentiable software that brings photon-number quantum optics and structured-light propagation into a common description, allowing the spatial evolution of light to directly influence non-classical quantum states and their interference. \textsc{esQueranto} is implemented in JAX, providing automatic differentiation and hardware-accelerated evaluation for optimization and automated search. We demonstrate the framework across a broad range of quantum-optical applications that exercise complementary aspects of its physical description. By combining quantum and structured-light physics within a single differentiable simulator, \textsc{esQueranto} takes an important step towards the dream of a foundational simulator in physics.
\end{abstract}
\maketitle

\section{Introduction}

Physics experiments are our way of asking questions to the Universe. Conventionally, these experiments have been designed by creative and experimenced human researchers. However, because the space of all possible experiments is enormously large, the question arises whether we have already discovered all ways to probe the Universe, or whether there are other high-performing but exceptionally unintuitive setups that human researchers have never (and potentially will never) envision.

This possibility motivates \textit{de-novo} experimental design, which uses advanced computational methods and artificial intelligence to explore the enormous search space of experiments, and discover new exceptional setups~\cite{klimesch2026designing}. A search algorithm chooses components, their connections, and their parameters, and uses a simulator to evaluate how well each setup meets a scientific objective. In quantum optics, this approach has produced new experimental designs for quantum state preparation, quantum measurements, and quantum-information tasks~\cite{krenn2016automated,knott2016search,ruiz2023pytheus,lanore2026automated} -- several of those have been implemented experimentally \cite{erhard2018experimental,wang2024entangling}. It allows us to explore experiments beyond the arrangements we would usually consider.

\begin{figure*}[!t]
    \centering
    \includegraphics[width=0.9\textwidth]{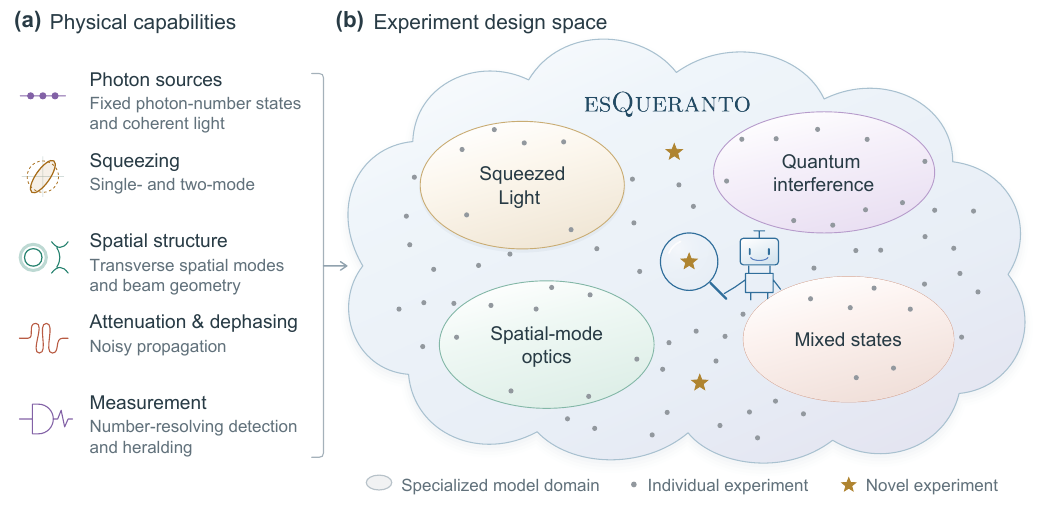}
    \caption{\textsc{esQueranto} combines physical capabilities (a) in a shared experiment design space (b). Gray dots represent individual experiments, and shaded regions indicate domains of specialized models. Stars highlight possible new experiments that combine capabilities beyond these domains; the robot illustrates automated search.}
    \label{fig:design-space}
\end{figure*}

The breadth of this exploration depends as much on the physical model used to evaluate an experiment as on the algorithm used to search for it. A richer physical description enlarges the space of experiments that can be explored, particularly when different processes can interact within the same model. Bringing atomic dynamics and structured quantum light into a common description, for example, makes it possible to search for sensors in which an atomic interaction reshapes a non-classical optical field and the resulting spatial change carries the signal. More generally, every additional physical capability, and every new coupling between capabilities, opens new directions for automated experimental design. The opportunity is therefore not simply to assemble a larger collection of components, but to describe diverse physical processes within a common framework in which their interactions can be followed consistently. This creates a shared design space in which combinations of physical effects become possible ingredients of experiments that may not have been considered before. This is the guiding idea behind \textsc{esQueranto}\footnote{\url{https://github.com/artificial-scientist-lab/esQueranto}}, the differentiable quantum-optical framework presented in this work, whose development is motivated by the connection between the physical processes we can model and the experiments we can discover in quantum optics (Fig.~\ref{fig:design-space}).

Our long-term vision is a universal simulator that brings different physical effects into a shared framework for experimental design~\cite{klimesch2026designing}. Existing software already provides powerful descriptions of complementary regions of this space. MrMustard, Piquasso, Perceval, and Strawberry Fields provide extensive tools for photonic quantum simulation~\cite{yao2024riemannian,kolarovszki2025piquasso,heurtel2023perceval,killoran2019strawberry}, while QuTiP provides a general framework for quantum dynamics with particularly broad support for open quantum systems~\cite{lambert2026qutip}. XLuminA provides a differentiable framework for scalar and vectorial wave optics, including diffraction, polarization and the automated design of complex optical and microscopy systems~\cite{rodriguez2024xlumina}. Chromatix is a JAX-based, GPU-accelerated differentiable wave-optics framework aimed at computational optics and microscopy, supporting scalar and vectorial propagation, polarization, diffractive optical elements, and propagation through complex three-dimensional scattering media~\cite{deb2026chromatix}. Finesse provides a detailed frequency-domain description of complex interferometers, combining coupled optical cavities and signal propagation with higher-order spatial modes, mode mismatch, quantum noise, optomechanical couplings, and control systems~\cite{freise2004finesse}. Differometor follows this frequency-domain interferometric description for carrier, signal, quantum-noise, and optomechanical fields, while introducing a JAX-based differentiable implementation designed for fast gradient-based optimization of large interferometers with many continuous parameters~\cite{klimesch2026differometor}. Table~\ref{tab:software-comparison} summarizes these complementary capabilities. The aim of \textsc{esQueranto} is not to replace these specialized descriptions, but to make new combinations of physical effects available within a common model.

\begin{table*}[t]
\centering
\caption{
Comparison of representative simulation frameworks across physical and
computational capabilities. Checkmarks indicate native support.
}
\label{tab:software-comparison}

\fontsize{8}{9.5}\selectfont
\renewcommand{\arraystretch}{1.2}
\setlength{\tabcolsep}{1.2pt}

\begin{tabular}{|>{\centering\arraybackslash\bfseries}c|c|c|c|c|c|c|c|c|c|c|}
\hline

&
\multirow{2}{*}{\textbf{\textsc{esQueranto}}}
& \multirow{2}{*}{\textbf{MrMustard}}
& \multirow{2}{*}{\textbf{Piquasso}}
& \multirow{2}{*}{\textbf{Perceval}}
& \textbf{Strawberry}
& \multirow{2}{*}{\textbf{QuTiP}}
& \multirow{2}{*}{\textbf{XLuminA}}
& \multirow{2}{*}{\textbf{Chromatix}}
& \multirow{2}{*}{\textbf{Finesse}}
& \multirow{2}{*}{\textbf{Differometor}}
\\
&
& & & &
\textbf{Fields}
& & & & &
\\
\hline

Fock
& \multirow{2}{*}{\cmark}
& \multirow{2}{*}{\cmark}
& \multirow{2}{*}{\cmark}
& \multirow{2}{*}{\cmark}
& \multirow{2}{*}{\cmark}
& \multirow{2}{*}{\cmark}
& \multirow{2}{*}{\xmark}
& \multirow{2}{*}{\xmark}
& \multirow{2}{*}{\xmark}
& \multirow{2}{*}{\xmark}
\\
representation
& & & & & & & & & & \\
\hline

Gaussian
& \multirow{2}{*}{\xmark}
& \multirow{2}{*}{\cmark}
& \multirow{2}{*}{\cmark}
& \multirow{2}{*}{\xmark}
& \multirow{2}{*}{\cmark}
& \multirow{2}{*}{\xmark}
& \multirow{2}{*}{\xmark}
& \multirow{2}{*}{\xmark}
& \multirow{2}{*}{\xmark}
& \multirow{2}{*}{\xmark}
\\
representation
& & & & & & & & & & \\
\hline

Photon\textsuperscript{a}
& \multirow{2}{*}{\cmark}
& \multirow{2}{*}{\cmark}
& \multirow{2}{*}{\cmark}
& \multirow{2}{*}{\cmark}
& \multirow{2}{*}{\cmark}
& \multirow{2}{*}{\cmark}
& \multirow{2}{*}{\cmark}
& \multirow{2}{*}{\cmark}
& \multirow{2}{*}{\cmark}
& \multirow{2}{*}{\cmark}
\\
loss
& & & & & & & & & & \\
\hline

\multirow{2}{*}{Dephasing}
& \multirow{2}{*}{\cmark}
& \multirow{2}{*}{\cmark}
& \multirow{2}{*}{\xmark}
& \multirow{2}{*}{\xmark}
& \multirow{2}{*}{\xmark}
& \multirow{2}{*}{\cmark}
& \multirow{2}{*}{\xmark}
& \multirow{2}{*}{\xmark}
& \multirow{2}{*}{\xmark}
& \multirow{2}{*}{\xmark}
\\
& & & & & & & & & & \\
\hline

Open-system
& \multirow{2}{*}{\xmark}
& \multirow{2}{*}{\xmark}
& \multirow{2}{*}{\xmark}
& \multirow{2}{*}{\xmark}
& \multirow{2}{*}{\xmark}
& \multirow{2}{*}{\cmark}
& \multirow{2}{*}{\xmark}
& \multirow{2}{*}{\xmark}
& \multirow{2}{*}{\xmark}
& \multirow{2}{*}{\xmark}
\\
dynamics
& & & & & & & & & & \\
\hline

Structured-light
& \multirow{2}{*}{\cmark}
& \multirow{2}{*}{\xmark}
& \multirow{2}{*}{\xmark}
& \multirow{2}{*}{\xmark}
& \multirow{2}{*}{\xmark}
& \multirow{2}{*}{\xmark}
& \multirow{2}{*}{\cmark}
& \multirow{2}{*}{\cmark}
& \multirow{2}{*}{\cmark}
& \multirow{2}{*}{\xmark}
\\
propagation
& & & & & & & & & & \\
\hline

Vectorial
& \multirow{3}{*}{\xmark}
& \multirow{3}{*}{\xmark}
& \multirow{3}{*}{\xmark}
& \multirow{3}{*}{\xmark}
& \multirow{3}{*}{\xmark}
& \multirow{3}{*}{\xmark}
& \multirow{3}{*}{\cmark}
& \multirow{3}{*}{\cmark}
& \multirow{3}{*}{\xmark}
& \multirow{3}{*}{\xmark}
\\
classical wave
& & & & & & & & & & \\
propagation
& & & & & & & & & & \\
\hline

Frequency
& \multirow{3}{*}{\xmark}
& \multirow{3}{*}{\xmark}
& \multirow{3}{*}{\xmark}
& \multirow{3}{*}{\xmark}
& \multirow{3}{*}{\xmark}
& \multirow{3}{*}{\xmark}
& \multirow{3}{*}{\xmark}
& \multirow{3}{*}{\xmark}
& \multirow{3}{*}{\cmark}
& \multirow{3}{*}{\cmark}
\\
domain
& & & & & & & & & & \\
interferometry
& & & & & & & & & & \\
\hline

Automatic
& \multirow{2}{*}{\cmark}
& \multirow{2}{*}{\cmark}
& \multirow{2}{*}{\cmark}
& \multirow{2}{*}{\xmark}
& \multirow{2}{*}{\cmark}
& \multirow{2}{*}{\cmark}
& \multirow{2}{*}{\cmark}
& \multirow{2}{*}{\cmark}
& \multirow{2}{*}{\xmark}
& \multirow{2}{*}{\cmark}
\\
differentiation
& & & & & & & & & & \\
\hline

\multirow{3}{*}{Focus}
& Structured
& Photonic
& Photonic
& Photonic
& Photonic
& Open
& Super
& Microscopy,
& GW
& GW
\\

& quantum
& quantum
& quantum
& quantum
& quantum
& quantum
& resolution
& holography
& detectors
& detectors
\\

& light
& optics
& computing
& computing
& computing
& systems
& microscopy
&
&
&
\\
\hline

\end{tabular}

\vspace{2pt}

\parbox{\textwidth}{
\fontsize{7}{8}\selectfont
\textsuperscript{a}Photon loss is represented differently across frameworks:
as a quantum loss channel in photon-number or quantum-state simulators,
as field attenuation or absorption in classical wave-optics frameworks,
and through lossy optical transfer in frequency-domain interferometer models.
}

\end{table*}

\textsc{esQueranto} expands the photonic quantum optics description with spatial beam geometry (such as Laguerre-Gauss) and light propagation. At a first sight, it might seem trivial to combine Fock-state quantum optics with spatial modes. After all, spatial modes form orthogonal bases, thus just expand the Fock-space as photons are just excitation of modes derived from Maxwell's equation. However, the situation is much more complex. Beam propagation and lenses change the spatial modes, thus change the basis of the Fock space. This can lead to partially orthogonal, partially non-orthogonal beams. Thus photons that started out in orthogonal modes could become partially indistinguishable during the propagation in the quantum experiment, which directly impacts the results of quantum interference. Furthermore, the spatial mode propagation introduces Gouy phases and accumulated Gouy phases \cite{erden1997accumulated, arai2013accumulated} with severe physical consequences \cite{zhou2017sorting,gu2018gouy}. Combining these Gaussian beam properties with the Fock space representation of quantum physics allows the simulation of complex, newly discovered phenomena such as the quantum Gouy phase \cite{hiekkamaki2022observation}.

This physical description is made directly accessible to automated experimental design through differentiability. \textsc{esQueranto} is implemented in JAX~\cite{bradbury2018jax}, providing automatic differentiation together with just-in-time compilation, vectorized and parallel execution, and hardware acceleration. Experimentally meaningful objectives can therefore be differentiated with respect to continuous source, optical, geometric, and noise parameters, enabling gradient-based optimization within a fixed experimental topology. The same evaluator can also be embedded within an external search over alternative experimental structures. \textsc{esQueranto} has already been used in automated-discovery studies of quantum experiments, including schemes that exploit higher-order pair emission to increase entanglement-generation success probabilities~\cite{ruiz2026automated} and the discovery of state-of-the-art heralding schemes for multiphoton path-entangled states~\cite{armezzani2026automated}.

We demonstrate the framework across six applications spanning different physical effects and experimental settings. We first consider multiphoton interference between nonlinear pair sources~\cite{herzog1994frustrated,hochrainer2022quantum,gu2019quantum,qian2023multiphoton,feng2023chip,gu2025analytical}, followed by the generation of NOON-like states~\cite{afek2010high} and the heralding of a path-entangled state~\cite{armezzani2026automated}. We then turn to high-precision interferometry with quantum states in the presence of loss and dephasing and to Hong--Ou--Mandel (HOM) interference between photons with structured transverse profiles, extending the description from photon-number dynamics to spatial-mode-dependent interference. Finally, we combine squeezed-light heralding, structured propagation and dephasing within a single setup, to perform radial-mode sorting~\cite{zhou2017sorting,gu2018gouy}. Together, these examples span state preparation, multiphoton interference, conditional measurements, noise and decoherence, parameter estimation, and spatial degrees of freedom, while the final example shows how placing these ingredients within a common description allows them to be composed into experiments that would otherwise belong to separate modeling domains.

\section{The \textsc{esQueranto} framework}
\label{sec:framework}

\noindent \textsc{esQueranto} provides a common representation for constructing and simulating quantum-optical experiments. A calculation follows the workflow

$
\texttt{space}
\rightarrow
\texttt{state}
\rightarrow
\texttt{circuit}
\rightarrow
\texttt{simulate}
\rightarrow
\texttt{measure}.
$

These objects describe distinct parts of the physical experiment: the space defines the Hilbert space in which the calculation is performed, the state specifies its quantum input, the circuit describes the sequence of physical transformations acting on it, and measurements extract probabilities, observables, or conditional output states. A minimal example is

\begin{pythoncode}
import esQueranto as eq
import jax.numpy as jnp

space = eq.space(2, cutoff=2, truncation="fixed")
initial = eq.state(space, (1, 1))

circuit = eq.circuit(
eq.bs(0, 1, jnp.pi / 4)
)

output = eq.simulate(initial, circuit)
probabilities = eq.pnr(output)
\end{pythoncode}

\noindent The same interface is retained as the physical description is extended to include coherent and squeezed sources, heralded state preparation, open-system propagation, and structured light, allowing these ingredients to be combined within a single experiment.

\subsection{State spaces and quantum states}
\label{sec:states}

\noindent \textsc{esQueranto} represents optical states in the Fock basis,
\begin{equation}
\ket{\mathbf n}
=
\ket{n_1,n_2,\ldots,n_M},
\end{equation}
where \(n_j\) is the photon occupation of mode \(j\). A pure state is therefore described by
\begin{equation}
\ket{\psi}
=
\sum_{\mathbf n} c_{\mathbf n}\ket{\mathbf n},
\end{equation}
while mixed states are represented by density operators in the same basis. Mixed states need not always be stored as full density matrices. When the state is block diagonal in total photon number, \textsc{esQueranto} can instead retain an exact factorized representation,
\begin{equation}
    \rho
    =
    \sum_n \sum_r
    w_{n,r}
    \ket{\psi_{n,r}}\!\bra{\psi_{n,r}},
\end{equation}
where each \(\ket{\psi_{n,r}}\) belongs to a definite total-photon-number sector. This representation is mathematically equivalent to the corresponding density matrix but can require substantially less storage when only a small number of factors is needed. \textsc{esQueranto} switches between ket, factorized, and dense representations internally as required by the evolution, without changing the user-level state interface. This provides a common quantum description for experiments ranging from fixed multiphoton inputs to coherent and squeezed states, noisy propagation, and conditional measurements.

The corresponding Hilbert space is created with \texttt{eq.space}, which specifies the number of modes $M$ and the photon-number cutoff retained in the calculation. Three truncation schemes are supported. With a \texttt{global} cutoff \(N\), all occupations satisfying
\begin{equation}
\sum_{j=1}^{M} n_j \leq N
\end{equation}
are retained, giving
\begin{equation}
D_{\mathrm{global}}=\binom{M+N}{N}.
\end{equation}
A \texttt{fixed} space contains only the sector
\begin{equation}
\sum_{j=1}^{M} n_j=N,
\end{equation}
with dimension
\begin{equation}
D_{\mathrm{fixed}}=\binom{M+N-1}{N},
\end{equation}
and is useful when the simulated evolution conserves the total photon number. Finally, a \texttt{mode} cutoff restricts each occupation independently to \(0\leq n_j\leq N\), resulting in \(D_{\mathrm{mode}}=(N+1)^M\). The appropriate truncation, with \texttt{global} as the default, therefore depends on the physical processes present in the experiment: fixed-number spaces are suitable for photon-number-conserving dynamics, whereas photon-generating or lossy processes require access to multiple photon-number sectors.

States are prepared with \texttt{eq.state}. A tuple specifies a single Fock state, while a mapping from occupations to complex amplitudes specifies an arbitrary superposition. For example,
\begin{pythoncode}
space = eq.space(2, cutoff=1)
psi = eq.state(
space,
{(1, 0): 1, (0, 1): 1j}
)
\end{pythoncode}
represents
\begin{equation}
\ket{\psi}
=
\frac{\ket{1,0}+i\ket{0,1}}{\sqrt{2}}.
\end{equation}
Complete ket and density-matrix representations can also be supplied directly, so that pure and mixed inputs are treated through the same state interface. States are normalized by default, while explicit control over normalization allows the probability weight associated with truncation or conditioning to be retained when required.

For optimization, \texttt{eq.state\_template} defines a fixed set of Fock occupations whose complex coefficients are treated as continuous parameters. This makes the input state itself differentiable while leaving its support, and hence the underlying Hilbert space, unchanged. The same state representation is also used when additional spatial degrees of freedom are introduced. Transverse structure simply enlarges the set of modes while preserving the Fock description of their photon occupations.

\subsection{Optical circuits and propagation}
\label{sec:optical-circuits}

Once an input state has been defined, its evolution in \textsc{esQueranto} is specified by an ordered sequence of optical elements. These elements are combined with \texttt{eq.circuit}, while \texttt{eq.simulate} propagates the state through the resulting circuit. For example,
\begin{pythoncode}
circuit = eq.circuit(
eq.bs(0, 1, jnp.pi / 4),
eq.ps(0, 0.3),
eq.bs(0, 1, jnp.pi / 4),
)

output = eq.simulate(initial, circuit)
\end{pythoncode}
defines a Mach-Zehnder interferometer independently of the input state. A circuit therefore specifies the physical topology of the experiment, while the input state and continuous parameters of the experiment can be varied separately.

The basic passive optical elements include beam splitters and phase shifters. A beam-splitter transformation between two modes is represented by \texttt{eq.bs}. For modes \(a\) and \(b\), \textsc{esQueranto} uses the convention
\begin{equation}
\label{eq:bs}
a^\dagger
\longrightarrow
ta^\dagger+rb^\dagger,
\qquad
b^\dagger
\longrightarrow
-r^*a^\dagger+tb^\dagger,
\end{equation}
where $t=\cos\theta$ and $r=e^{i\phi}\sin\theta$ are the transmission and reflection amplitudes, respectively. Accordingly, \texttt{eq.bs(a, b, theta, phi)} specifies the mixing angle \(\theta\) and reflection phase \(\phi\). A single-mode phase shift is represented by \texttt{eq.ps} and described by the unitary transformation
\begin{equation}
U_{\mathrm{ps}}(\phi)
=
e^{i\phi \hat n},
\end{equation}
where \(\hat n\) is the photon-number operator, so that an \(n\)-photon component acquires the phase \(e^{in\phi}\).  

\textsc{esQueranto} also supports coherent-state preparation through \texttt{eq.laser}, which prepares a coherent state in an initially empty mode. For a complex amplitude \(\alpha = A e^{i\phi}\), the prepared coherent state is
\begin{equation}
\ket{\alpha}
=
e^{-|\alpha|^2/2}
\sum_{n=0}^{\infty}
\frac{\alpha^n}{\sqrt{n!}}\ket{n},
\end{equation}
with mean photon number \(\langle\hat n\rangle=|\alpha|^2\) before truncation. 

Single-mode and two-mode squeezing are implemented through \texttt{eq.sms} and
\texttt{eq.tms}, respectively, with the squeezing operators
\begin{equation}
    S(r,\phi)
    =
    \exp\left[
        \frac{r}{2}
        \left(
            e^{-i\phi}a^2
            -
            e^{i\phi}a^{\dagger 2}
        \right)
    \right]
\end{equation}
and
\begin{equation}
    S_2(r,\phi)
    =
    \exp\left[
        r
        \left(
            e^{-i\phi}ab
            -
            e^{i\phi}a^\dagger b^\dagger
        \right)
    \right].
\end{equation}

For two-mode squeezing, the action on an arbitrary Fock
state can be written as
~\cite{gu2025analytical,chizhov1993photon,hu2008two,jabbour2021multiparticle},
\begin{align}
    S_2(r,\phi)\ket{p,q}
    &=
    \sum_{n=0}^{\min(p,q)}
    \sum_{k=0}^{\infty}
    \mathcal{A}_{n,k}^{(p,q)}\ket{p-n+k,q-n+k}
    \label{eq:tms_fock}
\end{align}
with
\begin{align}
    \mathcal{A}_{n,k}^{(p,q)}
    &=
    \frac{
        \left(e^{-i\phi}\tanh r\right)^n
        \left(-e^{i\phi}\tanh r\right)^k
    }{
        (\cosh r)^{p+q-2n+1}
    }
    \nonumber\\
    &\quad\times
    \left[
        \binom{p}{n}
        \binom{q}{n}
        \binom{p-n+k}{k}
        \binom{q-n+k}{k}
    \right]^{1/2}.
    \label{eq:tms_coefficient}
\end{align}
This expression gives the Fock-state amplitudes at arbitrary squeezing strength without requiring a recurrence relation, while making the contributions from pair annihilation and pair creation explicit. In a truncated Fock space, only terms whose output occupations lie within the retained basis are included. Single-mode squeezing changes photon number by pairs and therefore preserves the parity of the affected mode, while two-mode squeezing creates or annihilates one photon in each mode, preserving their photon-number difference and the parity of the total photon number.

Noise processes are described by quantum channels acting on the same optical modes. Photon loss is introduced through \texttt{eq.lossy\_channel}, whose parameter $\eta\in[0,1]$ is the intensity transmission. The channel is equivalent to coupling the selected mode to an environmental vacuum mode through a beam splitter. Defining
\begin{equation}
    B_{n,l}(\eta)
    =
    \sqrt{\binom{n}{l}}\,
    \eta^{(n-l)/2}(1-\eta)^{l/2},
\end{equation}
the channel acts in the Fock basis as
\begin{equation}
    |n\rangle\langle m|
    \longrightarrow
    \sum_{l=0}^{\min(n,m)}
    B_{n,l}(\eta)B_{m,l}(\eta)
    |n-l\rangle\langle m-l|.
\end{equation}

\noindent Photon-number dephasing is represented by \texttt{eq.dephasing\_channel}. Its parameter $\xi$ specifies the retained coherence between adjacent photon-number components, and the channel acts as
\begin{equation}
\ket{n}\bra{m}
\longrightarrow
\xi^{(n-m)^2}
\ket{n}\bra{m}.
\label{eq:dephasing}
\end{equation}
Thus \(\xi=1\) leaves all coherences unchanged, while \(\xi=0\) removes coherences between distinct number states without changing their populations.

Circuits can be evaluated once with \texttt{eq.simulate}, or compiled into a reusable evaluator with \texttt{eq.compile}. Compilation is useful when the experimental topology remains fixed while one or more continuous physical parameters are varied. A parameter is exposed by replacing its numerical value with a name and assigning an initial value during compilation. For example,
\begin{pythoncode}
circuit = eq.circuit(
eq.bs(0, 1, "theta")
)

evaluate, parameters = eq.compile(
initial,
circuit,
parameters={"theta": 0.3},
)

output = evaluate(parameters)
\end{pythoncode}
after which \texttt{parameters["theta"]} can be varied without reconstructing the circuit. The same mechanism applies to source amplitudes, phases, squeezing strengths, loss and dephasing parameters, and the geometrical quantities introduced for structured propagation below. Reusing the same parameter name in different elements makes them depend on the same physical quantity.

The evaluator returned by \texttt{eq.compile} can also include an observable or objective function evaluated from the propagated state and remains differentiable with respect to the exposed continuous parameters. This provides a common interface for parameter sweeps, gradient-based optimization, and automated experimental design while keeping the experimental topology fixed.

\subsection{Structured light and spatial propagation}
\label{sec:structured}

To incorporate the transverse structure of light, each optical path can be expanded into Laguerre--Gaussian (LG) modes labelled by the azimuthal index \(\ell\) and radial index \(p\). The resulting quantum modes are indexed by \((j,\ell,p)\), where \(j\) denotes the optical path. The transverse spatial profile of each mode is specified independently of its photon occupation in the Fock basis.

An LG profile is introduced with \texttt{eq.lg}, which specifies $(p,\ell)$ together with the beam waist, wavelength, and longitudinal position. Its geometry is described by the complex beam parameter
\begin{equation}
    q(z)=z+i z_R,
    \qquad
    z_R=\frac{\pi w_0^2}{\lambda},
\end{equation}
or equivalently
\begin{equation}
    \frac{1}{q(z)}
    =
    \frac{1}{R(z)}
    -
    i\frac{\lambda}{\pi w(z)^2},
\end{equation}
where $w_0$ is the waist radius, $w(z)$ the beam radius, and $R(z)$ the wavefront curvature. A structured input can, for example, be prepared as
\begin{pythoncode}
space = eq.space(2, cutoff=1, truncation="fixed")

profile = eq.lg(
    p=0,
    ell=0,
    waist=1e-3,
    wavelength=810e-9,
)

initial = eq.state(
    space,
    (1, 0),
    profiles={0: profile},
    radial_modes=20,
)
\end{pythoncode}
where \texttt{radial\_modes} sets the number of radial modes retained in the transverse basis.

The spatial profiles are propagated within scalar paraxial optics using the ABCD formalism. For an optical system described by the ray-transfer matrix
\begin{equation}
    M=
    \begin{pmatrix}
        A & B\\
        C & D
    \end{pmatrix},
    \label{eq:abcd_mat}
\end{equation}
the complex beam parameter transforms as
\begin{equation}
    q'
    =
    \frac{Aq+B}{Cq+D}.
    \label{eq:abcd_law}
\end{equation}
Free-space propagation over a distance $d$ and a thin spherical lens with focal length $f$ are represented by
\begin{equation}
    M_{\mathrm{free}}(d)
    =
    \begin{pmatrix}
        1 & d\\
        0 & 1
    \end{pmatrix},
    \qquad
    M_{\mathrm{lens}}(f)
    =
    \begin{pmatrix}
        1 & 0\\
        -1/f & 1
    \end{pmatrix},
    \label{eq:lens_free_space}
\end{equation}
and are implemented through \texttt{eq.free\_space} and \texttt{eq.lens}, respectively. These transformations describe diffraction and focusing through the evolution of the beam radius and wavefront curvature.

The transverse evolution also carries the Gouy phase
$\zeta(z)=\arctan\left(\frac{z}{z_R}\right)$, which is a mode-dependent phase.  For an LG mode of order $s_{\ell p}=2p+|\ell|+1$, the corresponding Gouy phase is \(s_{\ell p}\,\zeta(z)\). 

The Gouy phase accumulated in space in terms of an ABCD transformation and the complex beam parameter is \cite{arai2013accumulated},



\begin{equation}
    \Delta\zeta_{\ell p}
    =
    -s_{\ell p}
    \arg\left(A+\frac{B}{q}\right).
\end{equation}
For a thin lens, $
\Delta\zeta_{\ell p} = 0$, while for free propagation over a distance $d$ from $q=z+i z_R$, this becomes
\begin{equation}
    \Delta\zeta_{\ell p}(d)
    =
    s_{\ell p}
    \left[
        \arctan\left(\frac{z+d}{z_R}\right)
        -
        \arctan\left(\frac{z}{z_R}\right)
    \right].
\end{equation}
Different transverse orders therefore accumulate different phases as the field propagates, and an $n$-photon occupation acquires $n$ times the corresponding single-photon phase. This photon-number dependence of the Gouy phase has been observed experimentally for two-photon states~\cite{hiekkamaki2022observation}. Spatially uniform carrier phases are kept separate from this transverse contribution and can be introduced explicitly through \texttt{eq.ps}.

A second consequence of spatial propagation appears when the beam parameters of two fields are different. Although LG modes defined with different values of $q$ describe the same transverse Hilbert space, they correspond to different bases within it. If $u_{\ell p}(q)$ denotes a normalized LG mode, the change from a reference $q_1$ to another reference $q_2$ is described by
\begin{equation}
    U^{(\ell)}_{p'p}(q_2,q_1)
    =
    \braket{
        u_{\ell p'}(q_2)
        |
        u_{\ell p}(q_1)
    }.
    \label{eq:lg_overlap}
\end{equation}
For a single photon this gives
\begin{equation}
    \ket{1_{\ell p};q_1}
    =
    \sum_{p'=0}^{\infty}
    U^{(\ell)}_{p'p}(q_2,q_1)
    \ket{1_{\ell p'};q_2},
\end{equation}
showing that a photon occupying a single radial mode in one beam reference generally appears as a superposition of radial modes when expressed in another. The LG overlaps can be evaluated analytically~\cite{jaouni2025tutorial}, and the corresponding transformation extends directly to multiphoton Fock states.

This change of basis is essential when differently propagated fields interfere. If two inputs reach a beam splitter with references $q_1\neq q_2$, \textsc{esQueranto} first re-expresses both fields in a common transverse basis with a reference $\bar q$ chosen symmetrically from the two incoming beams,
\begin{equation}
    a^\dagger_{\ell p;q_i}
    =
    \sum_{p'}
    U^{(\ell)}_{p'p}(\bar q,q_i)
    a^\dagger_{\ell p';\bar q},
    \qquad i=1,2,
    \label{eq:modal_decomp}
\end{equation}
and the beam-splitter transformation is then applied to modes expressed in this shared basis. Differences in beam width and wavefront curvature therefore modify the amplitudes participating in the interference itself, rather than being represented by an external visibility factor.

For the cylindrically symmetric free-space and spherical-lens transformations considered here, the azimuthal index $\ell$ is conserved, while radial components can mix when the same field is represented in a different beam reference. Mode-dependent Gouy phases can in turn be converted into path-dependent interference, providing the physical mechanism underlying radial-mode sorting~\cite{zhou2017sorting,gu2018gouy}.

Overall, the structured-light description places photon-number evolution and transverse beam dynamics within the same quantum calculation, allowing diffraction, wavefront curvature, spatial-mode overlap, and mode-dependent phases to influence interference and measurement outcomes without introducing a separate phenomenological spatial model. The present description assumes paraxial, coaxial fields at a common wavelength, with temporal, spectral, and polarization degrees of freedom outside the spatial model.

\subsection{Measurements, conditional states, and fidelity}
\label{sec:measurement}

Measurements connect the propagated quantum state to outcome probabilities, expectation values of observables, and conditional output states. In \textsc{esQueranto}, photon counting is implemented through \texttt{eq.pnr}. For a state $\rho$ and a measurement outcome $h$ described by the effect $E_h$, the probability of that outcome is given by the Born rule
\begin{equation}
    p_h
    =
    \operatorname{Tr}\left(E_h\rho\right).
\end{equation}
For ideal photon-number-resolving detection, the effect associated with detecting $k$ photons is
\begin{equation}
    E_k
    =
    \ket{k}\bra{k}.
\end{equation}
Thus
\begin{pythoncode}
probabilities = eq.pnr(output)
\end{pythoncode}
returns the photon-count probability distribution of the output state.

Detector inefficiency is modeled by assuming that each photon is independently transmitted to an otherwise ideal detector with probability \(\eta_{\mathrm d}\). The resulting effect associated with detecting \(k\) photons is
\begin{equation}
    E_k^{(\eta_{\mathrm d})}
    =
    \sum_{n=k}^{\infty}
    \binom{n}{k}
    \eta_{\mathrm d}^{k}
    (1-\eta_{\mathrm d})^{n-k}
    \ket{n}\bra{n}.
\end{equation}
Finite photon-number resolution is introduced through \texttt{resolution=$r$}, for which the outcomes $0,1,\ldots,r$ are resolved individually and all larger counts are grouped into a final outcome corresponding to at least $r+1$ detected photons. In particular, \texttt{resolution=0} implements a threshold detector, distinguishing vacuum from the detection of one or more photons. Detector properties enter the same probability calculation, for example
\begin{pythoncode}
probabilities = eq.pnr(output, resolution=0,
                       efficiency=0.9)
\end{pythoncode}

When only a subset of the modes is measured, the measurement also prepares a conditional state on the unmeasured modes. Let \(A\) denote the retained subsystem and \(B\) the measured subsystem. For an outcome \(h\) on \(B\), the corresponding unnormalized conditional state of \(A\) is
\begin{equation}
    \widetilde{\rho}_{A|h}
    =
    \operatorname{Tr}_B
    \left[
        (I_A\otimes E_h)\rho_{AB}
    \right],
\end{equation}
with probability
\begin{equation}
    p_h
    =
    \operatorname{Tr}
    \widetilde{\rho}_{A|h},
\end{equation}
and normalized conditional state
\begin{equation}
    \rho_{A|h}
    =
    \frac{\widetilde{\rho}_{A|h}}{p_h},
    \qquad
    p_h>0.
\end{equation}
For an ideal photon-number-resolved measurement on a pure state, this reduces to
\begin{equation}
    \ket{\widetilde{\psi}_{A|h}}
    =
    {}_B\langle h|\psi_{AB}\rangle.
\end{equation}
In \textsc{esQueranto}, a conditional state is obtained by specifying the measured modes and measurement outcome,
\begin{pythoncode}
conditional, probability = eq.pnr(
    output,
    measured_modes=(2, 3),
    outcome=(1, 1)
)
\end{pythoncode}
\noindent which returns the normalized conditional state together with the probability of the heralding event.

For structured states, photon counting refers to the total occupation of each physical path,
\begin{equation}
    N_j
    =
    \sum_{\ell,p}
    n_{j\ell p}.
\end{equation}
If the transverse structure is not resolved by the detector, occupations with the same total path count contribute to the same measurement outcome and the unresolved transverse information is traced out when a conditional state is formed. A collection profile may instead be specified to select a transverse mode before photon counting, connecting spatial-mode selection to the same measurement formalism.

Conditional-state preparation is naturally characterized by both the probability of the selected event and the quality of the resulting state. The latter can be evaluated with \texttt{eq.fidelity}, which computes
\begin{equation}
    F(\rho,\sigma)
    =
    \left[
        \operatorname{Tr}
        \sqrt{
            \sqrt{\rho}\,
            \sigma
            \sqrt{\rho}
        }
    \right]^2.
\end{equation}
The comparison is performed directly through
\begin{pythoncode}
F = eq.fidelity(output, target)
\end{pythoncode}
\noindent and, for structured states described in different beam references, the transverse basis transformations introduced in Sec.~\ref{sec:structured} are included in the comparison. 

\subsection{Differentiable simulation and optimization}
\label{sec:optimization}

The parameterized circuit representation introduced above also defines the computational structure used for optimization. Once an experiment has been specified, its discrete structure---including the state-space layout, mode assignments, circuit topology, and measurement configuration---is fixed, while the quantities declared as parameters remain numerical inputs to the simulation. From this static description, \textsc{esQueranto} prepares the occupation maps and contraction structures required by the different optical transformations only once.

The static preparation is carried out at the NumPy level, whereas the subsequent numerical evolution is expressed in JAX. State amplitudes and exposed physical parameters therefore remain part of the traced computation, while the combinatorial structure of the experiment does not need to be reconstructed as their values change. The same evaluator can consequently be reused across parameter sweeps and optimization steps, with JAX handling differentiation and vectorized evaluation of the resulting numerical calculation. The computational benefits of batching and hardware acceleration depend on the size of the simulated Hilbert space and the number of experiments evaluated simultaneously. Appendix~\ref{app:scaling} benchmarks this crossover between CPU and GPU execution for passive universal interferometers~\cite{clements2016optimal}.

The compiled evaluator can be used inside user-defined objective functions, allowing the simulated experiment to be connected to arbitrary differentiable quantities of interest. For example, a loss function that depends on the fidelity with a target state can be written as
\begin{pythoncode}
evaluate, parameters = eq.compile(
    initial, setup,
    parameters=parameters,
)

def loss(parameters):
    output = evaluate(parameters)
    return 1.0 - eq.fidelity(output, target)

value, gradient = jax.value_and_grad(loss)(parameters)
\end{pythoncode}
\noindent Here the loss is defined independently of the circuit description, while differentiation propagates through the complete simulation. Alternatively, a callable acting directly on the propagated state can be supplied through the optional \texttt{objective} argument of \texttt{eq.compile}, so the evaluator returns that quantity instead of the state itself.

This division between a fixed experimental structure and differentiable numerical parameters is also useful beyond a single local optimization. Different structures can be proposed externally, compiled into their corresponding evaluators, and then optimized through the same interface. The simulator thus provides a common differentiable representation of each candidate experiment while leaving the exploration of alternative experimental structures to the surrounding design procedure. A complete optimization example is given in Appendix~\ref{app:optimization}.

\section{Applications of \textsc{esQueranto}}
\noindent We now show how \textsc{esQueranto} can be used in practice. The examples below cover squeezing, state preparation, conditional measurement, noisy propagation, and structured light; the last combines them in a single setup.

\subsection{Multiphoton Interference}
\begin{figure}[ht]
    \centering
    \includegraphics[width=1\linewidth]{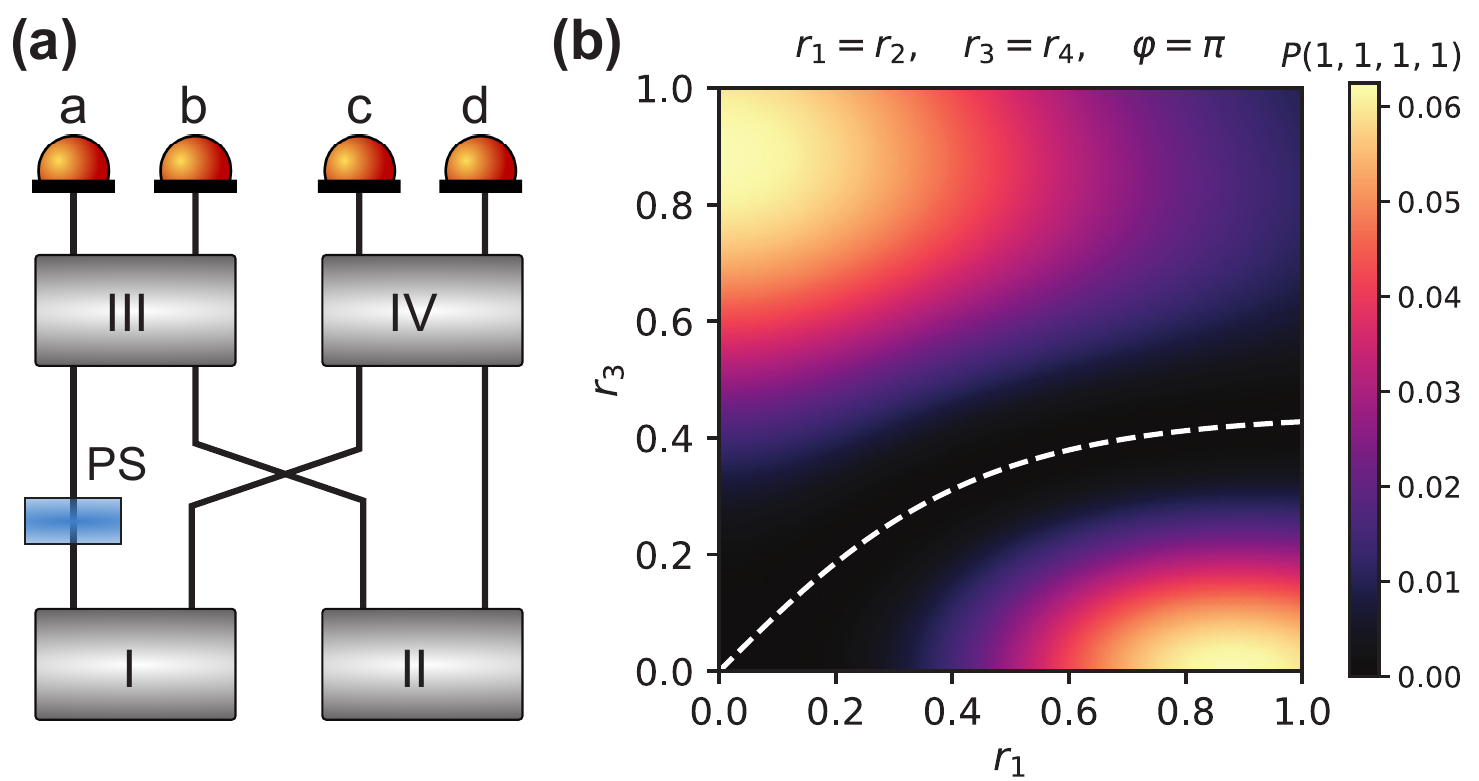}
    \caption{Four-crystal setup and four-photon coincidence probability. (a) Four two-mode squeezers act on modes $a,b,c,d$, with a phase shifter in path $a$ between sources I and III. (b) Simulated four-photon coincidence probability $P(1,1,1,1)$ as a function of $r_1$
    and $r_3$, with $r_1=r_2$, $r_3=r_4$, and $\phi=\pi$. The dashed line marks the analytical destructive-interference condition in Ref.~\cite{gu2025analytical}.}
    \label{fig:four-crystal}
\end{figure}

As a first example, we consider multiphoton interference between different
photon-generation pathways of a four-photon state. The configuration, shown in Fig.~\ref{fig:four-crystal}(a), consists of four two-mode squeezers acting on four optical modes $a,b,c,d$, with a phase shift $\phi$ in path $a$ between the first and second pair of squeezers. The corresponding circuit is
\begin{pythoncode}
setup = eq.circuit(
    eq.tms(0, 2, r1, 0.0),  # I: a,c
    eq.tms(1, 3, r2, 0.0),  # II: b,d
    eq.ps(0, phi),
    eq.tms(0, 1, r3, 0.0),  # III: a,b
    eq.tms(2, 3, r4, 0.0),  # IV: c,d
)
\end{pythoncode}

In the low-gain regime, a fourfold detection event $|1,1,1,1\rangle$ can
arise from two indistinguishable processes: sources I and II each emit one
photon pair, or sources III and IV each emit one pair. Because the phase
shifter acts only on the photon from source I in path $a$, the two amplitudes differ by a relative phase $\phi$ and cancel at $\phi=\pi$ when
$r_1=r_2=r_3=r_4$. This interference between photon-generation origins was
proposed in Ref.~\cite{gu2019quantum} and subsequently demonstrated in bulk and integrated photonic platforms~\cite{qian2023multiphoton,feng2023chip}. 

At stronger squeezing, however, this simple cancellation condition no longer holds. Higher-order pair-creation and annihilation processes provide additional contributions to the same output photon-number pattern, thereby modifying the interference condition. \textsc{esQueranto} includes these processes automatically by evolving the full Fock-space state rather than relying on a low-gain perturbative description. Fig.~\ref{fig:four-crystal}(b) shows the simulated four-photon probability $P(1,1,1,1)$ as a function of $r_1$ and $r_3$, with $r_1=r_2$, $r_3=r_4$, and $\phi=\pi$. As the squeezing increases, the position of the interference minimum shifts away from the low-gain condition because of the higher-order Fock components of the two-mode squeezed states, in agreement with the analytical high-gain treatment of Ref.~\cite{gu2025analytical}. This example shows how \textsc{esQueranto} captures interference effects beyond the low-gain regime.

\begin{figure}[!b]
    \centering
    \includegraphics[width=1\linewidth]{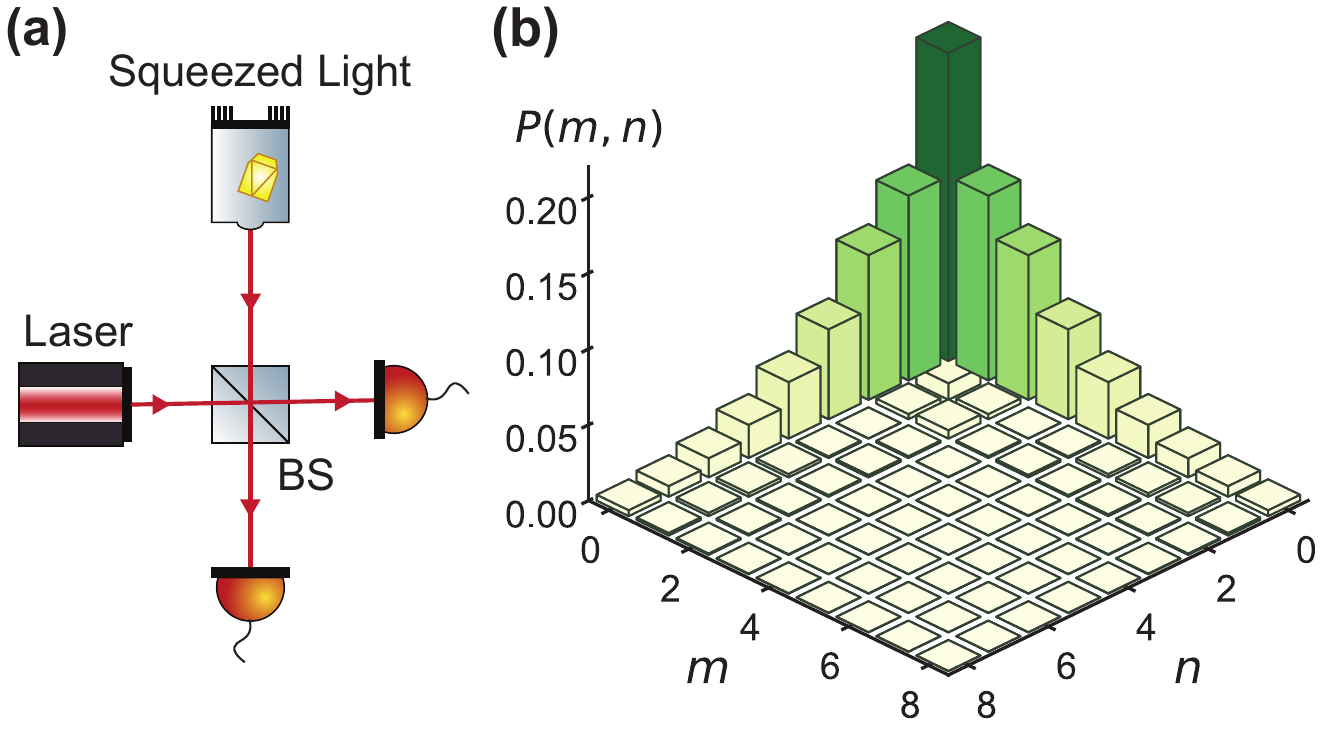}
    \caption{Generation of NOON-like states from coherent and squeezed light~\cite{afek2010high}. (a) A coherent field is mixed at a balanced beam splitter with a squeezed-vacuum state generated by collinear SPDC. (b) Simulated joint photon-number distribution \(P(m,n)\) at the two outputs, where the bar height gives the probability of detecting \(m\) and \(n\) photons. The larger bars near the two axes show that the photons tend to emerge together from either output port.}
    \label{fig:high-noon}
\end{figure}
\subsection{High-NOON states from classical and quantum light} 
NOON states are a prominent class of entangled states with applications in quantum sensing and metrology, where their N-fold phase response can enable phase super-resolution \cite{dowling2008quantum, mitchell2004super, slussarenko2017unconditional}. In a NOON state, all N photons occupy a coherent superposition of two optical paths. Such a state can be written as
\begin{equation}
\ket{\mathrm{NOON}_N}=\frac{\ket{N,0}+\ket{0,N}}{\sqrt{2}}.
\end{equation}

Generating NOON states at high photon numbers is experimentally challenging. One experimentally demonstrated approach is to interfere coherent light with squeezed vacuum at a balanced beam splitter~\cite{afek2010high}. The corresponding scheme is shown in Fig.~\ref{fig:high-noon}(a). A coherent state ($\ket{\alpha}$) is injected into one input port of the beam splitter, while a squeezed-vacuum state enters the other. The two inputs contain different photon-number components whose amplitudes interfere at the beam splitter. For suitable input amplitudes and phases, this interference suppresses intermediate output occupations within selected N-photon sectors while enhancing the $\ket{N,0}$ and $\ket{0,N}$ components, thereby producing NOON-like states.

In \textsc{esQueranto}, this configuration is specified directly using a coherent source, single-mode squeezing, and a balanced beam splitter:
\begin{pythoncode}
setup = eq.circuit(
    eq.laser(0, alpha, jnp.pi/2), # Coherent light 
    eq.sms(1, r, 0.0),   # Squeezed vacuum
    eq.bs(0, 1, jnp.pi/4, 0.0),  # 50:50 BS
)
\end{pythoncode}
The simulated joint photon-number distribution $P(m,n)$, where $m$ and $n$ denote the photon numbers in the two output modes, is shown in Fig.~\ref{fig:high-noon}(b).

\subsection{Heralding high-probability NOON states}
\label{sec:noon_heralding}

As an example of conditional state preparation, we revisit a high-probability heralding scheme discovered automatically with \textsc{esQueranto} in Ref.~\cite{armezzani2026automated}. Here, rather than repeating the discovery procedure, we use the resulting circuit to illustrate how a complete heralding experiment can be represented and evaluated within the framework.

We consider the optimized $\mathrm{NOON}_8$ scheme of Ref. \cite{armezzani2026automated}. The experiment starts from the four-mode input state $\ket{3,3,3,1}$, which is propagated through a compact interferometer consisting of five beam splitters. In \textsc{esQueranto}, the state and interferometer are specified as
\begin{pythoncode}
fock = eq.space(4, 10, truncation="fixed")
initial = eq.state(fock, (3, 3, 3, 1))

circuit = eq.circuit(
    eq.bs(0, 2, jnp.pi / 4, jnp.pi / 2),
    eq.bs(1, 3, jnp.pi / 8),
    eq.bs(0, 1, jnp.pi / 4),
    eq.bs(2, 3, jnp.pi / 4),
    eq.bs(1, 2, jnp.pi / 4)
)

propagate, parameters = eq.compile(initial, circuit)
output = propagate(parameters)
\end{pythoncode}

Two output modes are measured with photon-number-resolving detectors. Conditioning on the detection pattern $\ket{1,1}$ leaves eight photons in the two unmeasured modes and heralds the state
\begin{equation}
    \ket{\psi_{\mathrm{h}}}
    =
    \frac{\ket{8,0}-\ket{0,8}}{\sqrt{2}}.
\end{equation}
The corresponding conditional measurement and fidelity evaluation are implemented as
\begin{pythoncode}
noon8, success_probability = eq.pnr(
    output,
    measured_modes=(2, 3),
    outcome=(1, 1)
)
target = eq.state(
    noon8.space,
    {(8, 0): 1, (0, 8): -1}
)
fidelity = eq.fidelity(noon8, target)
\end{pythoncode}

\begin{figure}[h]
    \centering
    \includegraphics[width=\linewidth]{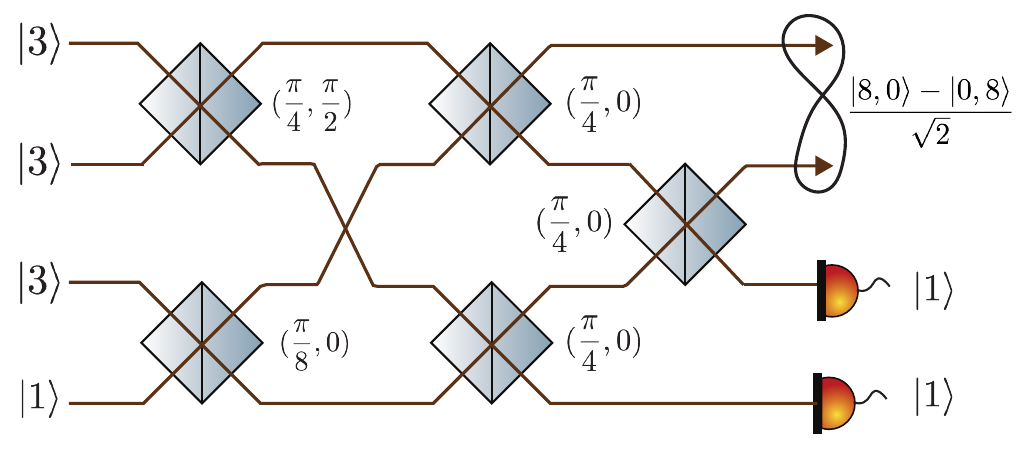}
    \caption{Heralded eight-photon path entanglement from the input $\ket{3,3,3,1}$, following Ref.~\cite{armezzani2026automated}.}
    \label{fig:heralded-noon}
\end{figure}

The circuit produces the target state with unit fidelity and improves the heralding probability by approximately $500\%$ relative to the best previously known human-designed scheme~\cite{zou2002generation}. The discovered scheme is shown in Fig.~\ref{fig:heralded-noon}.

\subsection{High-precision interferometry with loss and dephasing}

To demonstrate the treatment of photon loss and dephasing in \textsc{esQueranto}, we consider a high-precision interferometric sensing problem in which a quantum state accumulates a signal while both noise processes act during propagation. As a concrete example, we use the gravity-sensitive single-photon interferometer proposed in Ref.~\cite{zych2012general}, in which a photon is coherently split between two paths at different gravitational potentials. Recent long-baseline fiber experiments have brought this setting substantially closer to experimental realization~\cite{yvlk-t1j9}. The example therefore provides a physically motivated setting in which phase accumulation, photon loss, and dephasing must be treated simultaneously.

For this application, we introduce the \texttt{eq.fiber} element, which combines the phase accumulated during propagation with optional attenuation and dephasing. For a fiber of length \(L\) at height \(h\), the gravitational contribution to the optical phase is
\begin{equation}
    \phi_G
    =
    \frac{2\pi n_{\mathrm{r}} Lgh}
         {\lambda c^2},
    \label{eq:gravitational_shift}
\end{equation}
where \(n_{\mathrm{r}}\) is the refractive index, \(\lambda\) the wavelength, and \(g\) the gravitational acceleration. An \(n\)-photon component therefore acquires the phase \(e^{in\phi_G}\). Photon loss and dephasing use the same quantum channels introduced before. When their rates \(\alpha\) and \(\gamma\) are specified per kilometre, propagation over a distance \(L_{\mathrm{km}}\) gives
\begin{equation}
    \eta = e^{-2\alpha L_{\mathrm{km}}},
    \qquad
    \xi = e^{-\gamma L_{\mathrm{km}}},
\end{equation}
which determine the photon-loss and dephasing channels, respectively.

\begin{figure}[!t]
    \centering
    \includegraphics[width=0.96\linewidth]{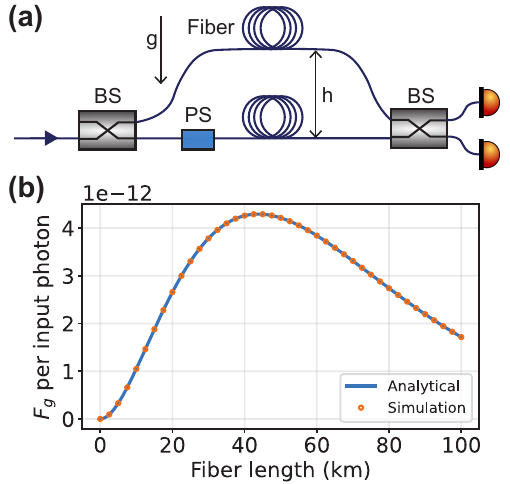}
    \caption{Gravity-sensitive fiber interferometer. (a) A single photon propagates in a superposition of two equal-length fiber paths separated vertically by $2\,\mathrm{m}$. (b) Fisher information for the gravitational acceleration $g$ as a function of the common fiber length at a phase bias close to quadrature.}
    \label{fig:MZIcombined}
\end{figure}

The interferometer is shown schematically in Fig.~\ref{fig:MZIcombined}(a). A single photon is prepared in one input mode and split by a balanced beam splitter into two equal-length fiber paths separated vertically by \(2\,\mathrm{m}\). In \textsc{esQueranto}, the single-photon input state and interferometer are described by
\begin{pythoncode}
fock = eq.space(2, 1, truncation="global")
initial = eq.state(fock, (1, 0))

topology = eq.circuit(
    eq.bs(0, 1, jnp.pi / 4),
    eq.ps(0, "phase_shift"),
    eq.fiber(
        0,
        length="fiber_length",
        height=0.0,
        g="gravity",
        attenuation=k_A,
        dephasing_rate=k_phi
    ),
    eq.fiber(
        1,
        length="fiber_length",
        height=2.0,
        g="gravity",
        attenuation=k_A,
        dephasing_rate=k_phi
    ),
    eq.bs(0, 1, jnp.pi / 4)
)
\end{pythoncode}

The height difference produces a relative gravitational phase that grows linearly with the common propagation length, while an independently adjustable phase shift sets the operating point of the interferometer. The output ports are measured with threshold detectors, and the resulting probabilities are differentiated with respect to \(g\) to obtain the Fisher information per incident photon.

Fig.~\ref{fig:MZIcombined}(b) shows the competition between signal accumulation and noise. Increasing the propagation length initially increases the gravitational phase and therefore the sensitivity to \(g\). At larger distances, however, photon loss reduces the detected signal while dephasing suppresses the interference visibility. Their competition produces an optimal length of approximately $44\,\mathrm{km}$. This example illustrates how \textsc{esQueranto} can evaluate quantum-interferometric sensing when the parameter-dependent evolution, photon loss, and dephasing all contribute to the final measurement statistics.

\subsection{HOM interference with structured photons}

Photons can exhibit complex spatial modes that capture non-trivial interference effects and interesting physics. The famed HOM interference effect describing the bunching of two indistinguishable photons can be altered depending on the photon's spatial profile, and its effects are observable depending on our detector measurement, which, itself, can admit spatial modes of a certain profile~\cite{jaouni2025tutorial}. Spatial propagation introduces an additional phase effect, the quantum Gouy phase, which can be exploited for mode-order sorting~\cite {gu2018gouy}, and is relevant to quantum sensing applications as well as to the observation of non-trivial modifications to the de Broglie wavelength~\cite {hiekkamaki2022observation, gu2022phase}. We therefore consider the generation, propagation, and interference of photons carrying structured spatial modes.
\begin{figure*}[ht]
    \centering
\includegraphics[width=1\textwidth]{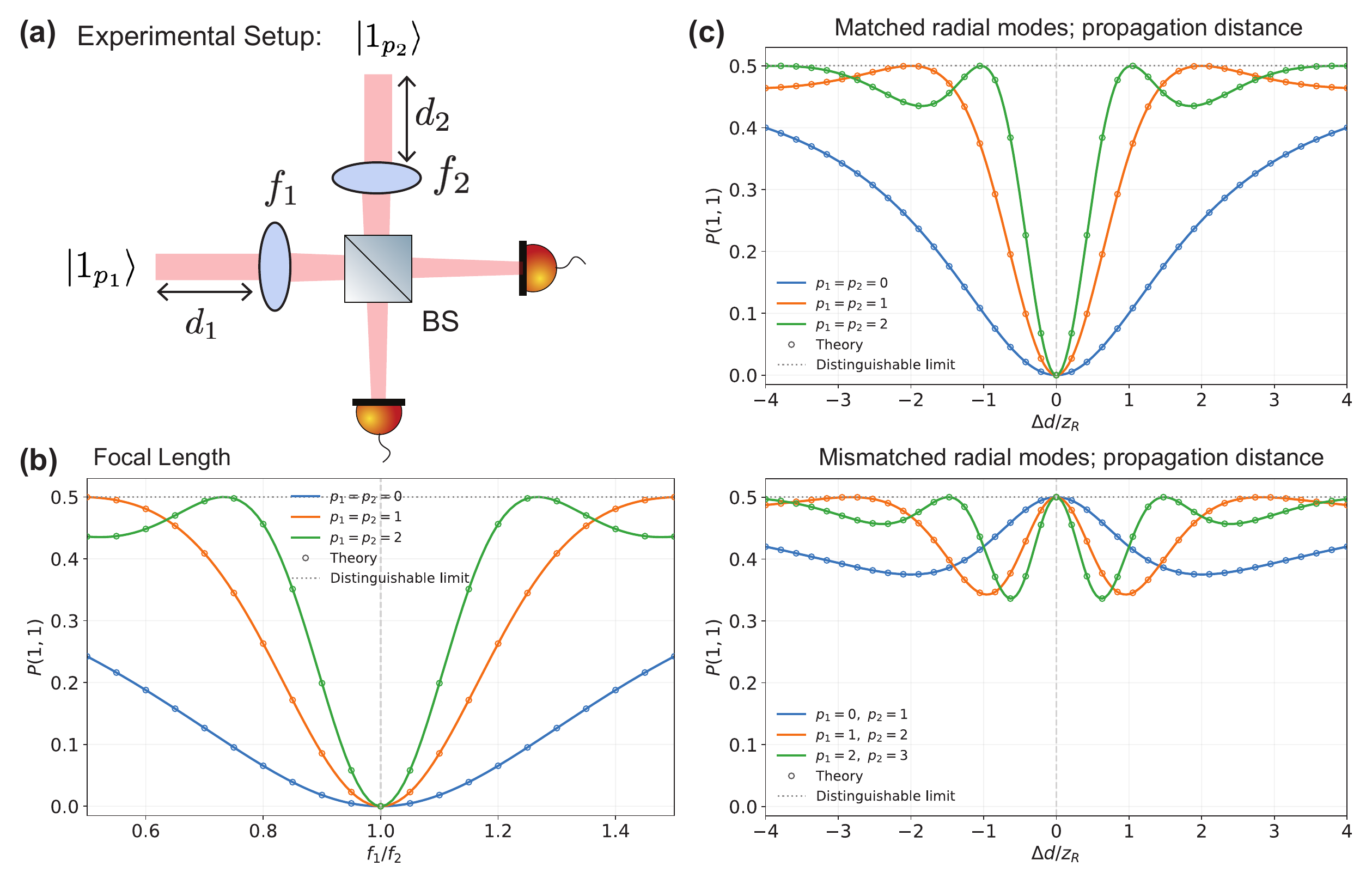}
    \caption{Hong--Ou--Mandel interference with structured photons. (a) One photon in each input path propagates through free space and a lens before a balanced beam splitter. (b) Coincidence probability versus focal-length ratio $f_1/f_2$. (c) Coincidence probability versus relative propagation distance $\Delta d/z_R$ for matched and mismatched radial modes.}
    \label{fig:structured_hom}
\end{figure*}

We illustrate how spatial modes can modulate the visibility of HOM interference. Fig.~\ref{fig:structured_hom} (a) shows the proposed experimental setup. The input state is
\begin{equation}
\ket{\psi_0}
=
a_{0p_1;q_{o}}^{\dagger} b_{0p_2;q_{o}}^{\dagger}\ket{0,0} = \ket{1_{p_1}, 1_{p_2}},
\end{equation}
with one photon in each input path. Both photons are initialized with the same complex beam parameter $q_0=i z_R$, but may occupy different radial modes $p_1$ and $p_2$. Along path $j$, the photon propagates over a distance $d_j$ and then passes through a lens of focal length $f_j$. Free-space propagation and lens transformation, as defined in Eq.~\eqref{eq:lens_free_space}, modify the complex beam parameter of each photon independently. As a result, the two photons can arrive at the beam splitter in different spatial bases. To interfere them within a common representation, both incoming states are decomposed into a shared basis defined by $\bar q$, following Eq.~\eqref{eq:modal_decomp}.
 
After the beam splitter, we evaluate the coincidence probability $P(1,1)$, corresponding to detecting one photon in each output port at the same time. We assume bucket detectors that do not resolve the transverse spatial mode. In this case, the HOM visibility is determined by the spatial overlap of the two incoming photons. For two arbitrary states described by $(p_1,q_1)$ and $(p_2,q_2)$, the coincidence probability is~\cite{jaouni2025tutorial}
\begin{equation}
    P(1,1) = \frac{1}{2}(1 - |U_{p',p}^{(0)}(q_{2}, q_{1})|^{2}),
\end{equation}
where $U_{p',p}^{(0)}(q_{2}, q_{1})$ is defined according to Eq.~\eqref{eq:lg_overlap}

The setup can be constructed in \textsc{esQueranto} as
\begin{pythoncode}
setup = eq.circuit(
    eq.free_space(0, d1),
    eq.free_space(1, d2), 
    eq.lens(0, f1),
    eq.lens(1, f2),
    eq.bs(0, 1, jnp.pi / 4, 0.0),
)
\end{pythoncode}

Fig.~\ref{fig:structured_hom}(b) and (c) show how the HOM coincidence dip changes as the optical geometry is varied. In Fig.~\ref{fig:structured_hom}(b), the focal lengths of the two lenses are varied, while in Fig.~\ref{fig:structured_hom}(c), the relative propagation distance is changed. Perfect HOM interference is recovered when the optical geometry produces spatially matched photons at the beam splitter. Away from this condition, the spatial overlap decreases and $P(1,1)$ approaches the distinguishable-photon limit of $1/2$. 

The dependence on geometry becomes more pronounced for higher radial-mode orders because their propagation acquires stronger mode-dependent Gouy phases. When the two photons occupy different radial modes, they remain orthogonal for matched beam parameters, giving the distinguishable-photon limit even when the optical geometries coincide. Changing the relative propagation distance alters the spatial bases of the two photons and can produce a finite overlap between the propagated modes, leading to a partial recovery of HOM interference.

\subsection{Heralded radial-mode sorting with dephasing}
\label{sec:combined-sorter}

\begin{figure}[!t]
    \centering
    \includegraphics[width=1\linewidth]{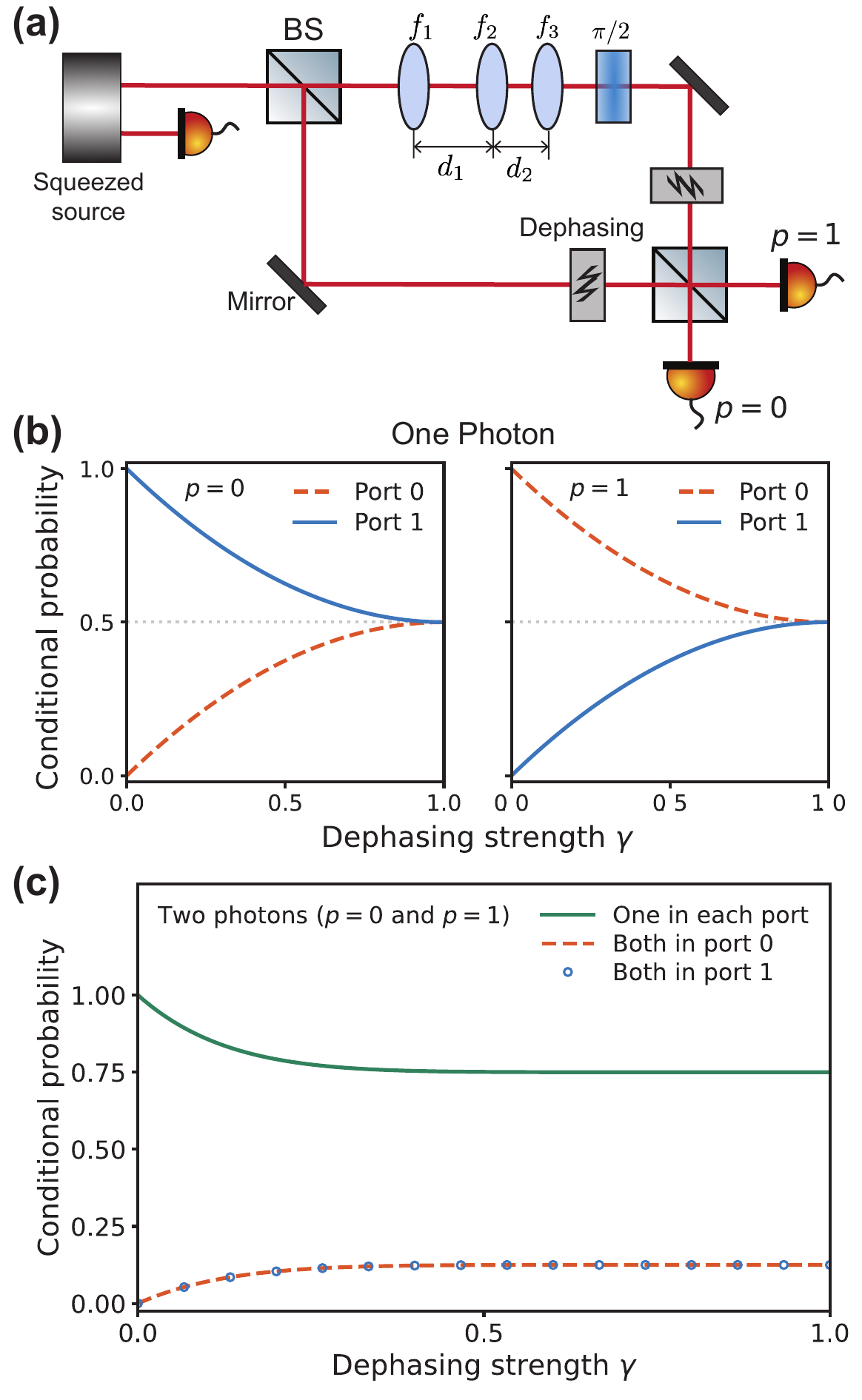}
    \caption{Radial-mode sorting with independent arm dephasing. The dephasing strength is $\gamma=1-\xi$, where $\xi$ is the one defined in Eq. \eqref{eq:dephasing}. (a) A single photon is heralded directly in a selected LG radial mode and sent through a three-lens interferometric sorter. (b) Output probabilities for $p=0$ and $p=1$, conditioned on successful heralding. (c) Output counts for a heralded two-photon input containing one photon in each radial mode in the same path.}
    \label{fig:combined-sorter}
\end{figure}

We finally combine squeezing, heralding, spatial-mode sorting, and noise in one experiment (Fig.~\ref{fig:combined-sorter}(a)). Two-mode squeezing acts directly on a selected Laguerre--Gaussian radial mode, $p=0$ or $p=1$, and detecting one photon in the ancillary mode heralds a single photon in the corresponding radial mode.

The sorter uses radial-mode-dependent Gouy phases to direct the two profiles toward opposite output ports~\cite{gu2018gouy}. Lenses and free propagation in a Mach--Zehnder interferometer produce a mode-dependent phase difference. We add dephasing noise in both arms before the second beam splitter, applied at the same propagation stage, and then count the output photons (without resolving their transverse modes, see Sec. \ref{sec:measurement}). The radial-mode sorter setup is described in \textsc{esQueranto} as:
\begin{pythoncode}
setup = eq.circuit(
    eq.bs(0, 1, jnp.pi / 4, -jnp.pi / 2),
    eq.lens(0, f1),
    eq.free_space(0, d1),
    eq.lens(0, f2),
    eq.free_space(0, d2),
    eq.lens(0, f3),
    eq.free_space(1, d1 + d2),
    eq.ps(0, jnp.pi / 2),
    eq.dephasing_channel(
        0, dephasing_fraction="coherence"
    ),
    eq.dephasing_channel(
        1, dephasing_fraction="coherence"
    ),
    eq.bs(0, 1, jnp.pi / 4, -jnp.pi / 2)
)
\end{pythoncode}

We first apply this setup to a heralded single photon in either the $p=0$ or $p=1$ radial mode. Fig.~\ref{fig:combined-sorter}(b) shows how dephasing progressively suppresses the sorting contrast. In the coherent limit, the two radial modes are directed deterministically to opposite output ports. As the dephasing strength increases, the interferometric distinction between them is reduced, and both inputs approach equal probabilities for the two outputs.

We then consider a two-photon state prepared by two independent two-mode squeezers acting on the $p=0$ and $p=1$ radial modes of the same signal path,
with separate ancillary modes. Conditioning on the detection of one photon in each ancillary mode directly heralds a signal containing one photon in $p=0$ and one in $p=1$. In the absence of dephasing, the two photons are routed to different ports with unit probability. Increasing the dephasing strength reduces this probability, but it approaches $3/4$ rather than $1/2$, while the two bunched outcomes each approach $1/8$, as shown in Fig.~\ref{fig:combined-sorter}(c). The residual correlation reflects the fact that the two radial components acquire the same random phase within a given interferometer arm, as shown explicitly by the two-photon calculation in Appendix \ref{app:explicit-p-sorter}.

\section{Outlook}

In this work, we introduced \textsc{esQueranto}, a differentiable framework for describing and optimizing quantum-optical experiments within a common computational language. By combining photon-number dynamics, squeezing, conditional measurements, mixed-state evolution, spatial propagation, and structured interference, we showed how physical ingredients that are often treated separately can be brought into the same simulation and connected to experimentally relevant objectives. The next step is to extend this description beyond quantum optics, with the long-term goal of building a foundational simulator in which increasingly different physical domains can be combined and optimized as parts of a single experiment.

Such a framework would open a broader design space in which quantum resources and signals can evolve across optical, mechanical, spatial, and many other degrees of freedom within one continuous description. A non-classical state could, for example, acquire spatial structure, couple to motion or matter, and ultimately be mapped onto a measurable observable, with the parameters governing each stage accessible to the same optimization. Concrete examples of how this broader framework can be developed are outlined below.

\textbf{Cavities and optomechanical couplings.}
Large interferometric systems introduce physical processes that are complementary to the present few-photon description. The frequency-domain treatment developed in Differometor~\cite{klimesch2026differometor}, for example, includes optical cavities, carrier and signal fields, quantum noise, and optomechanical response. Bringing such capabilities together with the Fock-space description of \textsc{esQueranto} would make it possible to investigate non-classical quantum resources inside realistic macroscopic interferometers. This could connect photon-resolved state preparation and measurement with resonant optical enhancement, mechanical motion, and frequency-dependent sensing within a common differentiable experiment.

\textbf{Cascaded Downconversion.} 
In \textsc{esQueranto} we model squeezing by assuming a classical pump. As a consequence, the quantum state can be described with Gaussian quantum optics. A genuine way outside of this picture is the idea of cascaded down-conversion \cite{hubel2010direct,hamel2014direct,krapick2016chip,kopf2025conservation}. Here, one pair creation process (such as squeezing) creates a large number of photon pairs. One of the two photon pairs is then used as the pump photon of another down-conversion process. Once succeeded, one has created three photons, that can be used for post-selection free quantum information protocols \cite{hamel2014direct} or the study of fundamental physical processes \cite{kopf2025conservation}. This phenomenon is also genuinely non-Gaussian and requires the full quantum description.

\textbf{Spatially resolved wave optics.}
The structured-light description of \textsc{esQueranto} could be extended toward the more general wave-optical capabilities available in XLuminA~\cite{rodriguez2024xlumina}. Beyond the present Laguerre--Gaussian and paraxial description, this could include arbitrary diffraction, phase and amplitude modulation, polarization, imaging systems, high-numerical-aperture optics, and vectorial fields. Combining these capabilities with photon-number quantum states would allow complex spatial transformations to directly influence multiphoton interference, state preparation, and quantum measurements. The spatial geometry of an experiment could then become a richer part of the same differentiable design space as its quantum-optical components.

\textbf{Matter-wave optics.}
A further direction is to extend the framework from photons to coherent massive particles, including electrons, ions, and molecular matter waves~\cite{hornberger2012quantum}. These systems introduce mass-dependent dispersion, matter-wave diffraction and interference, external electric and magnetic fields, and characteristic sources of decoherence. Describing optical and matter-wave elements within the same framework would make it possible to design experiments in which quantum information or sensing signals are transferred between different physical carriers. Their propagation, interactions, geometry, and environmental sensitivity could then become parts of the same differentiable experimental description.

\textbf{Atomic and many-body quantum systems.}
Internal atomic states and collective matter dynamics would provide another qualitatively different extension. Electronic, hyperfine, and spin degrees of freedom could introduce coherent light--matter interactions~\cite{hammerer2010quantum}, while ultracold gases and Bose--Einstein condensates could add interaction-driven nonlinear quantum dynamics. Combining these systems with non-classical and structured light would enable experiments in which optical states interact coherently with internal, motional, or collective matter degrees of freedom. 

Together, these directions chart the evolution of \textsc{esQueranto} toward a universal platform for experimental discovery, where complete experiments can be expressed within a single differentiable search space and new architectures can emerge from the interplay of their underlying physics, paving the way for automated discovery to become a foundational paradigm of experimental design.

\section*{Acknowledgments}
MK acknowledges support from the European Research Council (ERC) under the European Union's Horizon Europe research and innovation programme (ERC-2024-STG, Grant No.~101165179, ArtDisQ) and from the German Research Foundation (DFG; EXC 2064/1, Project 390727645). X.G. acknowledges support from the Alexander von Humboldt Foundation and the Deutsche Forschungsgemeinschaft (DFG, German Research Foundation), Project ID 398816777 (SFB 1375).
\bibliography{references}
\appendix

\clearpage
\newpage
\section{CPU--GPU scaling of batched optimization}
\label{app:scaling}

To quantify the computational scaling of the differentiable workflow introduced in Sec.~\ref{sec:optimization}, we benchmark the optimization of passive Clements interferometers~\cite{clements2016optimal}. We consider \(N\) photons in \(N\) modes, with \(N=4,\ldots,8\), and propagate the state in the fixed-\(N\) Fock space introduced in Sec.~\ref{sec:states}. 
For each value of \(N\), the circuit is a complete \(N\)-mode mesh containing \(N(N-1)/2\) beam splitters followed by \(N\) output phase shifters. Both the mixing angle and phase of every beam splitter are variable, so that the optimization contains \(N^2\) continuous parameters. 
For each benchmark, we first generate one random normalized \(N\)-photon input state. This state is kept fixed throughout the optimization:

\begin{pythoncode}
space = eq.space(N, N, truncation="fixed")
initial = eq.state(space, random_amplitudes)
\end{pythoncode}

A reachable target is generated by propagating the same input through another Clements interferometer with independently randomized parameters:

\begin{pythoncode}
target = eq.simulate(
    initial,
    random_target_circuit,
)
\end{pythoncode}

The benchmark therefore does not test whether the optimizer can reach an arbitrary state outside the chosen circuit family. Instead, it measures the cost of optimizing a known reachable transformation. The candidate circuit is compiled once,

\begin{pythoncode}
propagate, parameters = eq.compile(
    initial,
    setup,
    parameters=parameters,
)
\end{pythoncode}

and the loss is simply the infidelity with the target, using the fidelity introduced in Sec.~\ref{sec:measurement}:

\begin{pythoncode}
def loss(parameters):
    output = propagate(parameters)
    return 1.0 - eq.fidelity(output, target)
\end{pythoncode}

To expose parallelism, we optimize \(B\) independent initial parameter sets simultaneously. They share the same input and target but have independent circuit parameters and Adam optimizer states~\cite{kingma2015adam}. The loss and gradient of every member of the batch are evaluated together using \texttt{vmap}:

\begin{pythoncode}
loss_and_grad = jax.vmap(
    jax.value_and_grad(loss)
)

values, grads = loss_and_grad(parameters)
\end{pythoncode}

A single compiled optimization step then advances the complete batch:

\begin{pythoncode}
updates, opt_state = optimizer.update(
    grads, opt_state, parameters
)
parameters = optax.apply_updates(
    parameters, updates
)
\end{pythoncode}

To compare CPU and GPU execution, we run the same optimization on either eight Intel Xeon Platinum 8358 CPU cores or one NVIDIA A100-SXM-64GB GPU. We report the speedup factor

\begin{equation}
    \frac{t_{\mathrm{CPU},8}(N,B)}
         {t_{\mathrm{GPU}}(N,B)},
\end{equation}

where \(B\) is the number of optimization instances evaluated simultaneously. Values \(>1\) indicate faster GPU execution.

\begin{figure}[H]
    \centering
    \hspace*{-0.5cm}
    \includegraphics[width=0.95\linewidth]
        {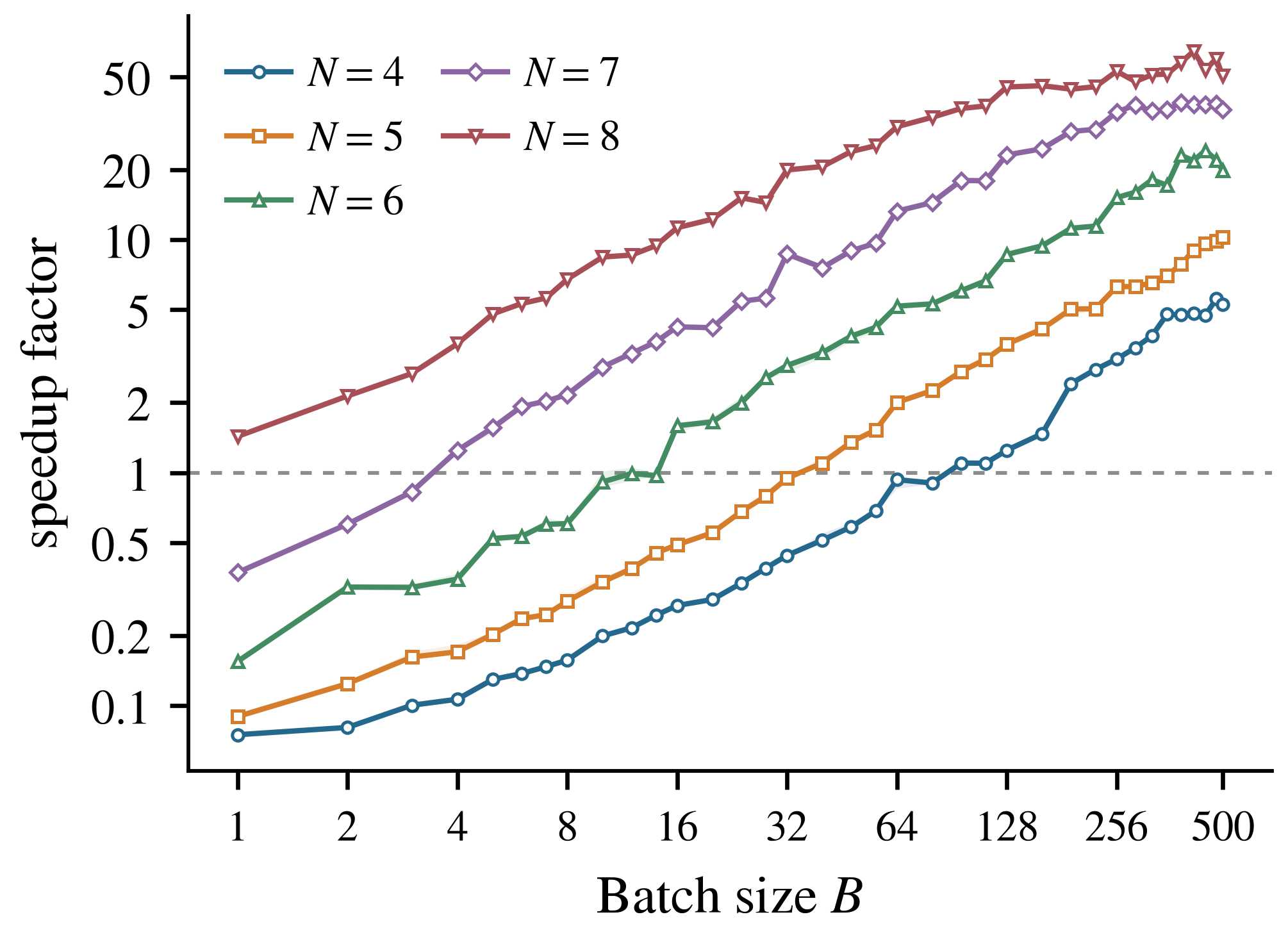}
    \caption{CPU--GPU scaling of batched optimization for Clements interferometers containing \(N=4,\ldots,8\) photons in \(N\) modes. The speedup factor compares eight Intel Xeon Platinum 8358 CPU cores with one NVIDIA A100-SXM-64GB GPU. The dashed line marks equal execution time.}
    \label{fig:cpu-gpu-scaling}
\end{figure}

Figure~\ref{fig:cpu-gpu-scaling} shows the transition from CPU- to GPU-favorable execution as the computational workload increases. For small photon numbers and small batches, the CPU is faster, whereas increasing either \(N\) or \(B\) progressively favors the GPU. Larger Fock spaces increase the amount of computation required for each optimization instance, while batching exposes additional parallelism. Consequently, the crossover to GPU acceleration occurs at increasingly smaller batch sizes as the simulated system grows.
\section{Example of gradient-based optimization}
\label{app:optimization}

As a concrete illustration of the differentiable optimization introduced in Sec.~\ref{sec:optimization}, we consider the $\mathrm{NOON}_8$ heralding problem of Sec.~\ref{sec:noon_heralding}. We start from the four-mode input
\begin{equation}
    \ket{\psi_{\mathrm{in}}}=\ket{3,3,3,1},
\end{equation}
which in \textsc{esQueranto} is defined as

\begin{pythoncode}
fock = eq.space(4, 10, truncation="fixed")
initial = eq.state(fock, (3, 3, 3, 1))
\end{pythoncode}

The target state on the two unmeasured modes is

\begin{equation}
    \ket{\psi_{\mathrm{t}}}
    =
    \frac{\ket{8,0}-\ket{0,8}}{\sqrt{2}},
\end{equation}

represented by

\begin{pythoncode}
target_space = eq.space(2, 8, truncation="fixed")
target = eq.state(
    target_space,
    {(8, 0): 1, (0, 8): -1},
)
\end{pythoncode}

We use a generic four-mode Clements interferometer~\cite{clements2016optimal} as initial ansatz:

\begin{pythoncode}
setup = eq.circuit(
    eq.bs(0, 1, "theta0", "phi0"),
    eq.bs(2, 3, "theta1", "phi1"),
    eq.bs(1, 2, "theta2", "phi2"),
    eq.bs(0, 1, "theta3", "phi3"),
    eq.bs(2, 3, "theta4", "phi4"),
    eq.bs(1, 2, "theta5", "phi5"),
    eq.ps(0, "alpha0"),
    eq.ps(1, "alpha1"),
    eq.ps(2, "alpha2"),
    eq.ps(3, "alpha3"),
)
\end{pythoncode}

All continuous beam-splitter and phase parameters are initialized independently from uniform distributions over their respective domains. Denoting these initial values by \texttt{parameters}, the circuit is compiled once into a differentiable evaluator:

\begin{pythoncode}
propagate, parameters = eq.compile(
    initial,
    setup,
    parameters=parameters,
)
\end{pythoncode}

The two final modes are measured with photon-number-resolving detection, and the accepted heralding event is
\begin{equation}
    h=\ket{1,1}.
\end{equation}
For each parameter set, \textsc{esQueranto} therefore returns both the conditional state and its heralding probability. We optimize the scalar loss
\begin{equation}
    \mathcal{L}
    =
    (1-F)
    +
    \lambda(1-p_h),
    \label{eq:noon8-optimization-loss}
\end{equation}
where \(F\) is the fidelity of the heralded state with the target and \(p_h\) is the probability of the heralding event. The parameter \(\lambda\) controls the relative weight assigned to success probability. The corresponding objective is

\begin{pythoncode}
def loss(parameters):
    output = propagate(parameters)

    heralded, p_h = eq.pnr(
        output,
        measured_modes=(2, 3),
        outcome=(1, 1),
    )

    F = eq.fidelity(heralded, target)
    return (1-F) + lam*(1-p_h)
\end{pythoncode}

The continuous parameters are then optimized directly with Adam:

\begin{pythoncode}
optimizer = optax.adam(learning_rate)
opt_state = optimizer.init(parameters)

for _ in range(4000):
    value, grad = jax.value_and_grad(loss)(parameters)
    updates, opt_state = optimizer.update(
        grad, opt_state, parameters
    )
    parameters = optax.apply_updates(parameters, updates)
\end{pythoncode}

\begin{figure}[t]
    \centering
    \includegraphics[width=0.78\linewidth]
    {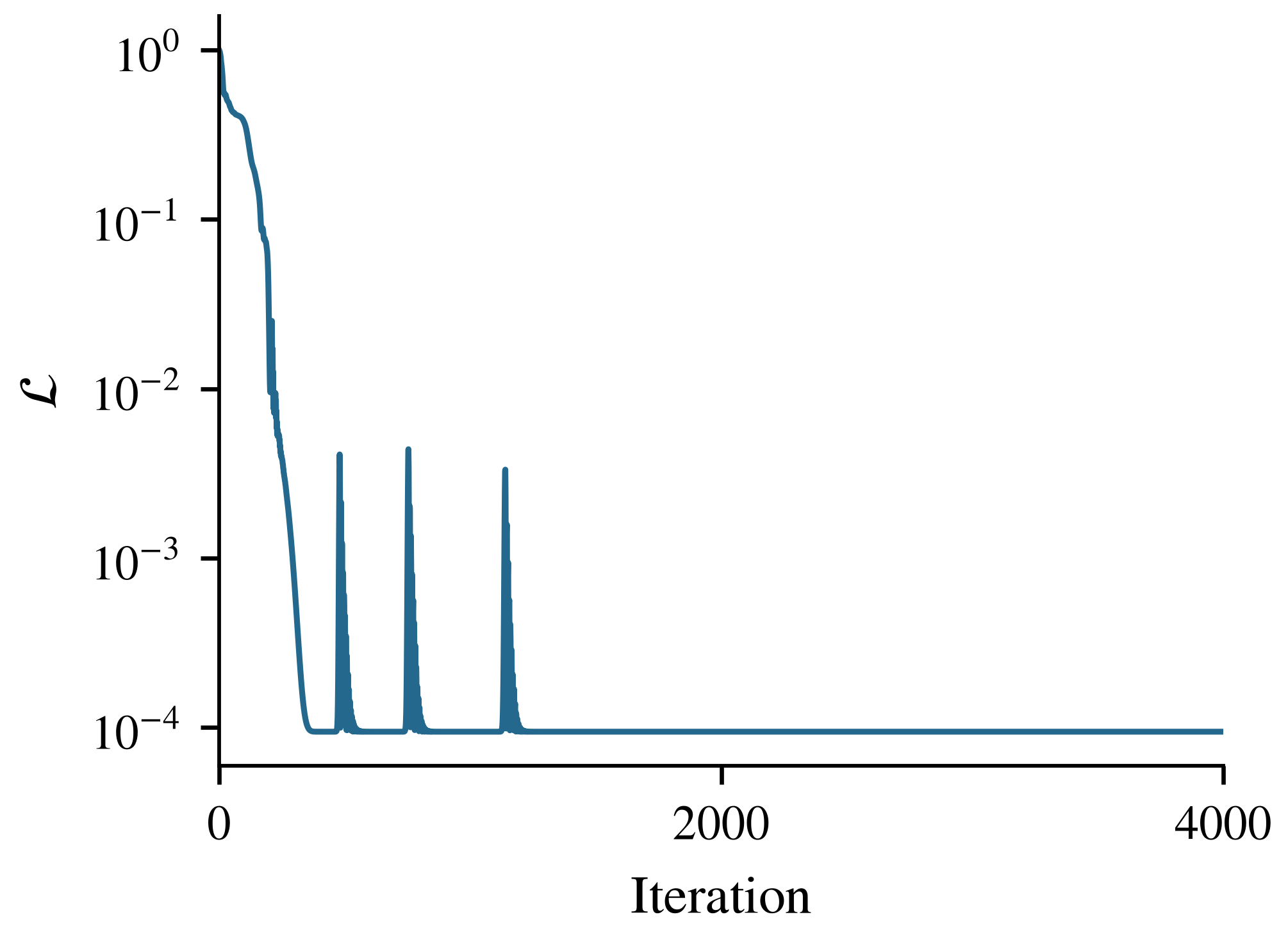}
    \caption{Loss $\mathcal{L}$ during a representative optimization of the $\mathrm{NOON}_8$ heralding problem from a randomly initialized four-mode Clements interferometer.}
    \label{fig:noon8-loss}
\end{figure}

Figure~\ref{fig:noon8-loss} shows the evolution of the objective during a representative optimization. Starting from the randomly initialized Clements interferometer, the loss decreases by approximately four orders of magnitude and rapidly reaches its minimum. The optimization was performed on an NVIDIA A100 GPU; 4000 Adam steps required \(18.8\,\mathrm{s}\), corresponding to an average of approximately \(4.7\,\mathrm{ms}\) per optimization step.

\begin{figure}[t]
    \centering
    \includegraphics[width=0.80\linewidth]
    {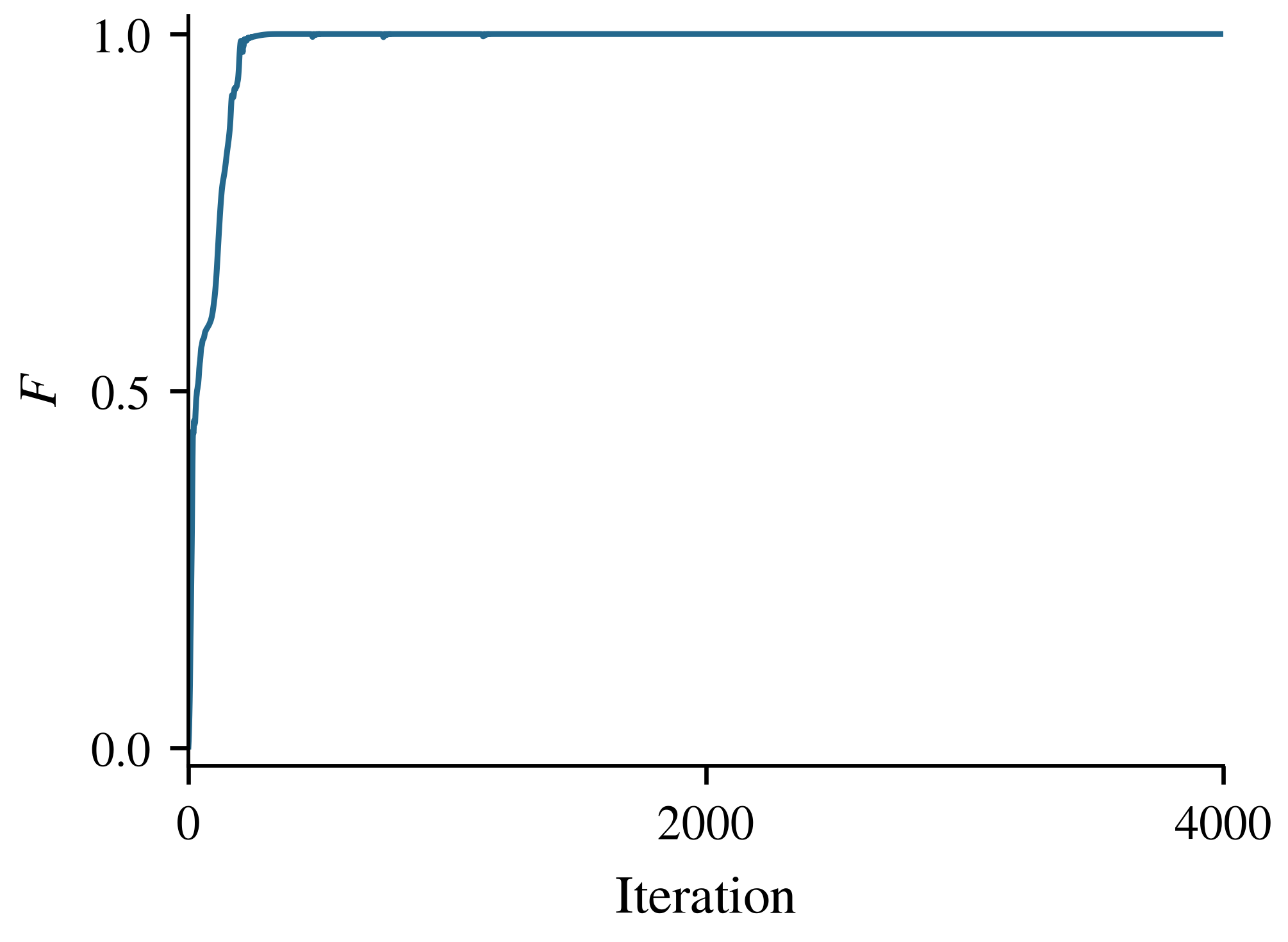}

    \vspace{0.2em}

    \includegraphics[width=0.80\linewidth]
    {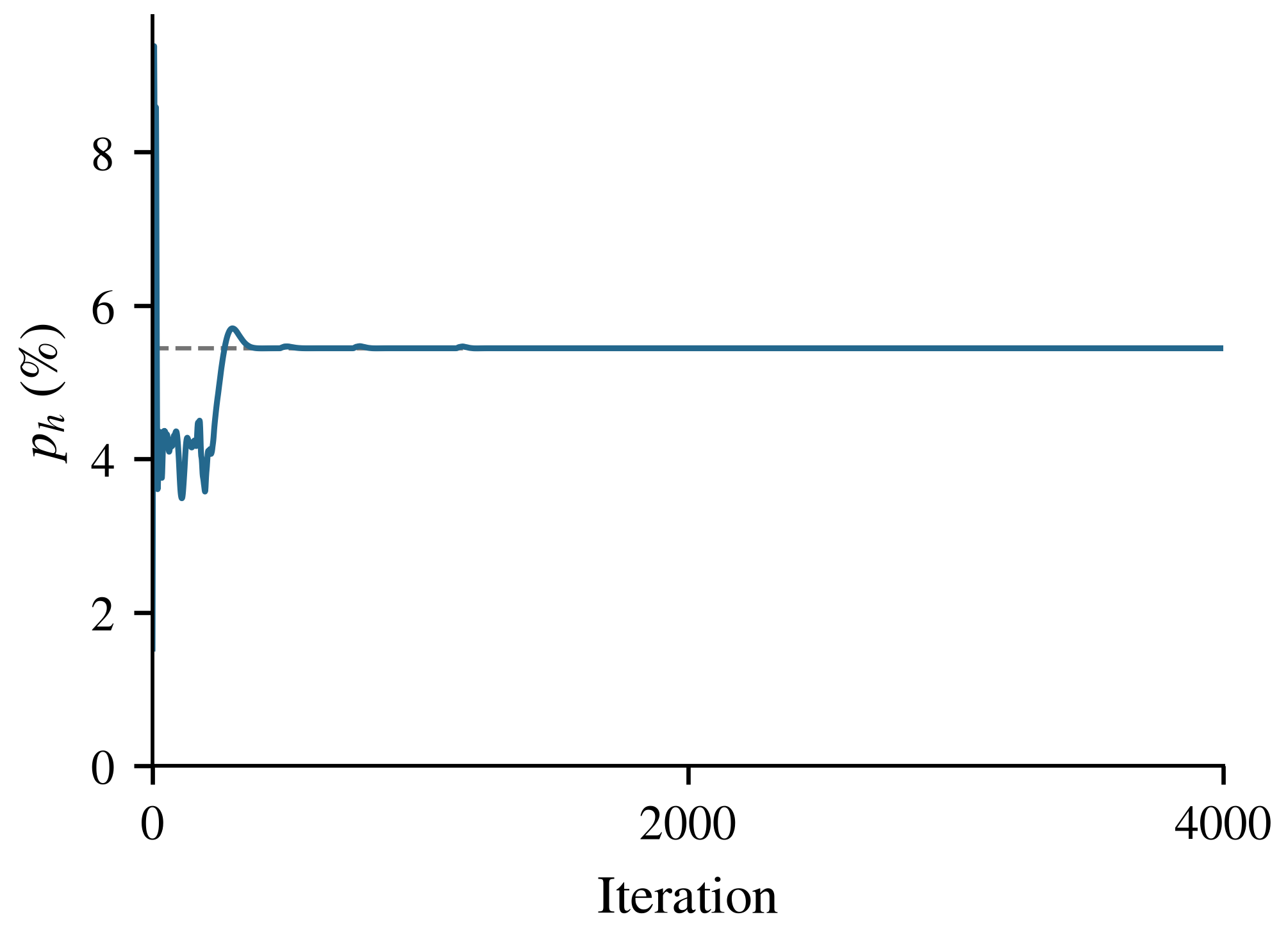}

    \caption{Quantities contributing to the optimization objective in Fig.~\ref{fig:noon8-loss}. Top: fidelity $F$ of the heralded state with the target $\mathrm{NOON}_8$ state. Bottom: heralding probability $p_h$ for the outcome $h=\ket{1,1}$.}
    \label{fig:noon8-components}
\end{figure}

The two terms entering the loss reveal how this convergence is achieved. As shown in Fig.~\ref{fig:noon8-components}, the fidelity initially rises rapidly and approaches unity, establishing the desired heralded state. At the same time, the optimization adjusts the interferometer toward a solution with a heralding probability of approximately \(5.4\%\), corresponding to the high-probability $\mathrm{NOON}_8$ scheme discussed in Sec.~\ref{sec:noon_heralding}. The small excursions visible during the optimization reflect the simultaneous optimization of state fidelity and heralding probability through the combined objective.

\section{Radial-mode sorter under dephasing}
\label{app:explicit-p-sorter}

Here we derive the dephasing dependence of the single- and two-photon probabilities shown in Fig.~\ref{fig:combined-sorter}. Throughout this experiment $\ell=0$, and only the radial modes $p=0$ and $p=1$ are populated. Conditioning a mode-selective two-mode squeezed source on one photon in its ancillary mode prepares a single photon in the selected radial mode, while conditioning two independent sources prepares
\begin{equation}
    a^\dagger_{00;q}a^\dagger_{01;q}|0\rangle ,
\end{equation}
with one photon in each radial mode of the same input path. 

At the final beam splitter the two arms are expressed in the same transverse reference $q$. For the balanced beam splitters used in the sorter, $\theta=\pi/4$ and $\phi=-\pi/2$, so the convention of Eq.~\ref{eq:bs} gives
\begin{equation}
    a^\dagger_{\ell p;q}
    \longrightarrow
    \frac{
        a^\dagger_{\ell p;q}
        -i b^\dagger_{\ell p;q}
    }{\sqrt{2}},
    \qquad
    b^\dagger_{\ell p;q}
    \longrightarrow
    \frac{
        -i a^\dagger_{\ell p;q}
        +b^\dagger_{\ell p;q}
    }{\sqrt{2}}.
\end{equation}
The three-lens arm is chosen such that the relative phase accumulated by $p=1$ differs by $\pi$ from that of $p=0$. Up to an overall phase, the states immediately before the dephasing channels and final beam splitter are consequently
\begin{align}
    p=0:\qquad
    |\psi\rangle
    &=
    \frac{
        a^\dagger_{00;q}
        -i b^\dagger_{00;q}
    }{\sqrt{2}}
    |0\rangle ,
    \\
    p=1:\qquad
    |\psi\rangle
    &=
    \frac{
        -a^\dagger_{01;q}
        -i b^\dagger_{01;q}
    }{\sqrt{2}}
    |0\rangle .
\end{align}

The two interferometer arms undergo independent number dephasing with the same retained coherence $\xi$. Since the channel acts on the total photon occupation of each physical path, its action on a path coherence is
\begin{equation}
    |n_a,n_b\rangle\langle m_a,m_b|
    \longrightarrow
    \xi^{(n_a-m_a)^2+(n_b-m_b)^2}
    |n_a,n_b\rangle\langle m_a,m_b| .
\end{equation}
For a single photon, the coherence between the two arms therefore acquires the factor $\xi^2$. Applying the final balanced beam splitter gives
\begin{align}
    p=0:\qquad
    P(1,0)
    &=
    \frac{1-\xi^2}{2},
    &
    P(0,1)
    &=
    \frac{1+\xi^2}{2},
    \\
    p=1:\qquad
    P(1,0)
    &=
    \frac{1+\xi^2}{2},
    &
    P(0,1)
    &=
    \frac{1-\xi^2}{2}.
\end{align}
Thus $\xi=1$ gives deterministic routing to opposite ports, whereas $\xi=0$ gives equal probabilities for the two outputs, reproducing Fig.~\ref{fig:combined-sorter}(b).

For the two-photon input, the first balanced beam splitter and the radial-mode-dependent phase produce, immediately before dephasing,
\begin{equation}
\begin{split}
    |\psi\rangle
    =
    \frac{1}{2}
    \big(
        &-a^\dagger_{00;q}a^\dagger_{01;q}
        -i a^\dagger_{00;q}b^\dagger_{01;q}
        +i b^\dagger_{00;q}a^\dagger_{01;q}
        -b^\dagger_{00;q}b^\dagger_{01;q}
    \big)|0\rangle .
\end{split}
\end{equation}
The four terms have total path occupations $(2,0)$, $(1,1)$, $(1,1)$, and $(0,2)$, respectively. Their coherences are therefore affected differently by the path dephasing. Coherence between the $(2,0)$ and $(0,2)$ components is multiplied by
\begin{equation}
    \xi^{(2-0)^2+(0-2)^2}
    =
    \xi^8,
\end{equation}
whereas coherences connecting either of these components to a $(1,1)$ component acquire $\xi^2$. Crucially, the coherence between
\begin{equation}
    a^\dagger_{00;q}b^\dagger_{01;q}|0\rangle
    \qquad\text{and}\qquad
    b^\dagger_{00;q}a^\dagger_{01;q}|0\rangle
\end{equation}
is unaffected by dephasing, because both components contain exactly one photon in each physical path. The channel resolves path photon number, not the radial-mode assignment within a path.

Applying the final balanced beam splitter and summing over the unresolved radial modes gives
\begin{align}
    P(2,0)
    &=
    P(0,2)
    =
    \frac{1-\xi^8}{8},
    \\
    P(1,1)
    &=
    \frac{3}{4}
    +
    \frac{\xi^8}{4}.
\end{align}
The terms proportional to $\xi^2$ cancel when the two possible radial-mode assignments contributing to the same path-count outcome are combined. For $\xi=1$, the photons are therefore routed to different output ports with unit probability. At complete dephasing, $\xi=0$,
\begin{equation}
    P(1,1)=\frac{3}{4},
    \qquad
    P(2,0)=P(0,2)=\frac{1}{8},
\end{equation}
which explains the nonclassical limiting values in Fig.~\ref{fig:combined-sorter}(c).

\end{document}